\documentclass[1p,preprint,10pt]{elsarticle}

\usepackage{amsmath}
\usepackage{amsfonts}
\usepackage{array}
\usepackage{xcolor}
\usepackage{xspace}
\usepackage{hyperref}
\usepackage{lineno}
\usepackage{listings}
\usepackage{tabularx}
\usepackage{tabulary}
\usepackage{makecell}
\usepackage{placeins}

\newcommand{\TRIQS}{\textnormal{\texttt{TRIQS}}\xspace}
\newcommand{\CTHYB}{\textnormal{\texttt{CTHYB}}\xspace}
\newcommand{\SOM}{\textnormal{\texttt{SOM}}\xspace}

\DeclareMathOperator{\atan}{atan}
\DeclareMathOperator{\sign}{sign}
\newcommand{\dilog}{\ensuremath{\mathrm{Li}_2}}

\graphicspath{{./figs/}}

\definecolor{darkblue}{rgb}{0,0,.6}
\definecolor{darkred}{rgb}{.6,0,0}
\definecolor{darkgreen}{rgb}{0,.6,0}
\definecolor{red}{rgb}{.98,0,0}

\def\ssmall{\fontsize{8pt}{2pt}\selectfont}

\lstnewenvironment{pylisting}[1][]
{\lstset{language=Python,numbers=none,#1}}{}
\newcommand\py[1]{\lstinline[language=Python]{#1}}

\journal{Computer Physics Communications}

\begin{document}

\begin{frontmatter}

\title{\TRIQS/\SOM: Implementation of the Stochastic Optimization Method for Analytic Continuation}

\author[Hamburg]{Igor Krivenko\corref{author}}
\ead{ikrivenk@physnet.uni-hamburg.de}
\author[Hamburg]{Malte Harland}
\ead{mharland@physnet.uni-hamburg.de}

\cortext[author] {Corresponding author. Current address: Department of Physics,
University of Michigan, Ann Arbor, Michigan 48109, USA}

\address[Hamburg]{I. Institut f\"ur Theoretische Physik, Uni. Hamburg, Jungiusstra\ss e 9, 20355 Hamburg, Germany}

\begin{abstract}
    We present the \TRIQS/\SOM analytic continuation package, an efficient
    implementation of the Stochastic Optimization Method proposed by A.
    Mishchenko {\it et al} [Phys. Rev. B {\bf 62}, 6317 (2000)].
    \TRIQS/\SOM strives to provide a high quality open source
    (distributed under the GNU General Public License version 3)
    alternative to the more widely adopted Maximum Entropy continuation
    programs. It supports a variety of analytic continuation problems
    encountered in the field of computational condensed matter physics.
    Those problems can be formulated in terms of response functions of
    imaginary time, Matsubara frequencies or in the Legendre polynomial basis
    representation. The application is based on the \TRIQS C++/Python
    framework, which allows for easy interoperability with \TRIQS-based quantum
    impurity solvers, electronic band  structure codes and visualization tools.
    Similar to other \TRIQS packages, it comes with a convenient Python
    interface.
\end{abstract}

\end{frontmatter}

\noindent {\bf PROGRAM SUMMARY}

\begin{small}
\noindent
{\em Program Title:} \TRIQS/\SOM \\
{\em Project homepage:} \url{http://krivenko.github.io/som/} \\
{\em Program Files doi:} \url{http://dx.doi.org/10.17632/fcjzyhrwpw.1} \\
{\em Journal Reference:} \url{http://dx.doi.org/10.1016/j.cpc.2019.01.021} \\
{\em Operating system:} Unix, Linux, macOS \\
{\em Programming language:} \verb*#C++#/\verb*#Python#\\
{\em Computers:} any architecture with suitable compilers including PCs and clusters \\
{\em RAM:} Highly problem-dependent \\
{\em Distribution format:} GitHub, downloadable as zip \\
{\em Licensing provisions:} GNU General Public License (GPLv3) \\
{\em Classification:} 4.8, 4.9, 4.13, 7.3 
{\em External routines/libraries:}
  \verb#TRIQS 1.4.2#\cite{TRIQS}, \verb#Boost >=1.58#, \verb#cmake#.\\
{\em Nature of problem:}
Quantum Monte Carlo methods (QMC) are powerful numerical techniques widely used
to study quantum many-body effects and electronic structure of correlated
materials. Obtaining physically relevant spectral functions from noisy QMC \
results in the imaginary time/Matsubara frequency domains
requires solution of an ill-posed analytic continuation problem as a post-processing step.\\
{\em Solution method:}
We present an efficient C++/Python open-source implementation of the stochastic
optimization method for analytic continuation.
\end{small}

\section{\label{sec:introduction}Introduction}

Quantum Monte Carlo methods (QMC)\cite{Gull11RMP} are one of the most
versatile classes of numerical tools used in quantum many-body
theory and adjacent fields. When applied in combination with the dynamical mean 
field theory (DMFT)\cite{VollhardtMetzner,DMFT,DMFTReview} and an electronic 
structure calculation method, such as the density functional theory (DFT), they 
also allow to assess diverse effects of electronic correlations in real 
materials\cite{Kotliar2006}.

In spite of a considerable progress in development of real time quantum Monte
Carlo solvers \cite{Muehlbacher08,Werner09,RealTimeCTAUX,Chen17A,Chen17B},
QMC methods formulated in the imaginary time
\cite{Rubtsov2005,Werner2006,Gull2008,Otsuki2007,Prokofev2007,Prokofev2008,
Blankenbecler1981} remain a more practical choice in most situations. Such
algorithms have a fundamental limitation as they calculate dynamical response
functions in the imaginary time/Matsubara frequency representation, or in a
basis of orthogonal polynomials in more recent
implementations\cite{Boehnke2011}.
Getting access to the spectral functions that are directly observable in the
experiment requires solution of the infamous analytic continuation problem. The
precise mathematical statement of the problem varies between response functions
in question and representations chosen to measure them. However, all 
formulations amount to the solution of an inhomogeneous Fredholm integral equation
of the first kind w.r.t. to the spectral function $A(\epsilon)$ and with 
an approximately known left hand side $O(\xi)$,
\begin{equation}\label{eq:fredholm_continuous}
O(\xi) = \int\limits_{\epsilon_\mathrm{min}}^{\epsilon_\mathrm{max}}
d\epsilon\ K(\xi,\epsilon) A(\epsilon).
\end{equation}

Such equations are known to be ill-posed by Hadamard, i.e. neither existence nor
uniqueness of their solutions is guaranteed \cite{Groetsch2007}. 
Moreover, behaviour of the solution is unstable w.r.t. the changes in the 
LHS. This last feature makes the numerical treatment of the problem 
particularly hard as the LHS is known only up to the stochastic QMC noise.
There is little hope that a general solution of the problem of numerical 
analytic continuation can be attained. Nonetheless, the ongoing effort to 
develop useful continuation tools has given rise to dozens of diverse methods.

One of the earliest continuation methods, which is still in use, is based on
construction of Pad\'e approximants of the input
data\cite{Vidberg1977,Zhu2002,Beach2000}. In general, its applicability is
strongly limited to the cases with high signal-to-noise ratio of the input.

Variations of the Maximum Entropy Method (MEM) currently represent the {\it de
facto} standard among tools for numerical analytic continuation
\cite{Jarrell1996,Bryan1990}. While not being computationally demanding, they
allow to obtain reasonable results for realistic input noise levels.
There are high quality open source implementations of MEM available to the 
community, such as \TRIQS/\texttt{maxent} \cite{Kraberger2017,TRIQSMaxent},
$\Omega$\textit{Maxent} \cite{Bergeron2016} together with its compatibility layer
\texttt{OmegaMaxEnt\_TRIQS} \cite{OmegaMaxEnt_TRIQS}, and a package
based on the ALPSCore library \cite{Levy2017}. The typical criticism of the MEM
targets its bias towards the specified default model \cite{Vafayi2007}.

Another group of approaches, collectively dubbed stochastic continuation 
methods, exploit the idea of random sampling of the ensemble of possible 
solutions (spectral functions). The method of A. Sandvik is one of the
most well-known members of this family \cite{Sandvik1998}. It can be shown
that the MEM is equivalent to the saddle-point approximation of the statistical
ensemble considered by Sandvik \cite{Beach2004}.

The Stochastic Optimization Method (SOM) \cite{Mishchenko2000,Goulko2017}, also
referred to as Mishchenko method, is a stochastic algorithm that demonstrates 
good performance in practical tests \cite{Schott2016}. It is bias-free in the
sense that it does not favour solutions close to a specific default model, or
those possessing special properties beyond the standard non-negativity
and normalization requirements. It also features a continuous-energy
representation of solutions, which is more natural than the traditional
projection on a fixed energy mesh. The price to pay for
SOM's advanced features is its relatively high CPU time demands.
In some cases, running SOM simulations may require as much time
as running a QMC solver for producing input data.

The lack of a high quality open source implementation of such a 
well-established and successful method urged us to write the program 
package called \SOM. Besides, until the very recent
releases of \TRIQS/\texttt{maxent} and \texttt{OmegaMaxEnt\_TRIQS}
there were no good analytic continuation tool compatible with otherwise
powerful Toolbox for Research on Interacting Quantum Systems (\TRIQS) \cite{TRIQS},
so we decided to fill this gap. (\TRIQS Green's function library features
only a simplistic and very limited implementation of the Pad\'e method).

There are other continuation methods that are worth mentioning. Among these are
the average spectrum method \cite{Syljuasen2008}, the method of consistent
constraints (MCC) \cite{Prokofev2013}, stochastic methods based on the Bayesian
inference \cite{Vafayi2007, Fuchs2010}, the Stochastic Optimization
with Consistent Constraints protocol (SOCC, a hybrid between SOM and MCC that
allows for characterization of the result accuracy) \cite{Goulko2017}, and
finally the recent and promising sparse modelling approach
\cite{Otsuki2017}.

This paper is organized as follows. In Section \ref{sec:som} we recapitulate
the Stochastic Optimization Method. Section \ref{sec:optimizations} gives a
brief description of the performance optimizations devised in our 
implementation of SOM. We proceed in Section \ref{sec:usage} with instruction on
how to use the program. Section \ref{sec:results} contains some illustrational
results of analytic continuation with \SOM. Practical information on how to
obtain and install the program is given in Section \ref{sec:starting}. Section
\ref{sec:summary} summarizes and concludes the paper.
Some technical details, which could be of interest to authors of similar codes, 
can be found in the appendices.

\section{\label{sec:som}Stochastic Optimization Method}

We will now briefly formulate the analytic continuation problem, and summarize
Mishchenko's algorithm \cite{Pavarini2012} as implemented in \SOM.

\subsection{\label{sec:problem}Formulation of problem}

Given a function $O(\xi)$ (referred to as \textit{observable} throughout the
text), whose numerical values are approximately known at $M$ discrete points
$\xi_m$, we seek to solve a Fredholm integral equation of the first kind,
\begin{equation}\label{eq:fredholm}
O(\xi_m) = \int\limits_{\epsilon_\mathrm{min}}^{\epsilon_\mathrm{max}}
d\epsilon\ K(\xi_m,\epsilon) A(\epsilon),
\end{equation}
w.r.t. spectral function $A(\epsilon)$ defined on interval
$[\epsilon_\mathrm{min};\epsilon_\mathrm{max}]$.
The observable $O$ is usually obtained as a result of QMC accumulation,
and therefore known with some uncertainty. It is also assumed that values
of $O(\xi_m)$ for different $m$ are uncorrelated.
The definition  domain of $A(\epsilon)$ as well as the explicit form of the 
integral kernel $K(\xi,\epsilon)$ vary between different kinds of observables and 
their representations (choices of $\xi$). The sought spectral function is required
to be smooth, non-negative within its domain, and is subject to
normalization condition
\begin{equation}\label{eq:normalization}
    \int\limits_{\epsilon_\mathrm{min}}^{\epsilon_\mathrm{max}}
    d\epsilon\ A(\epsilon) = \mathcal{N},
\end{equation}
where $\mathcal{N}$ is a known normalization constant.

Currently, \SOM supports four kinds of observables. Each of them can be given
as a function of imaginary time points on a regular grid $\tau_m = 
\beta\frac{m}{M-1}$ ($\beta$ is the inverse temperature), as Matsubara
components, or as coefficients of the Legendre polynomial expansion 
\cite{Boehnke2011}. This amounts to a total of 12 special cases of equation 
(\ref{eq:fredholm}). Here is a list of all supported observables.

\begin{enumerate}
    \item Finite temperature Green's function of fermions,
          $G_{\alpha\alpha}(\tau) = -\langle\mathbb{T}_\tau c_\alpha(\tau) c_\alpha^\dagger(0)\rangle$.
          
          $G(\tau)$ must fulfil the anti-periodicity condition
          $G(\tau+\beta) = -G(\tau)$. The off-diagonal elements
          $G_{\alpha\beta}(\tau)$ are not supported, since they are not
          in general representable in terms of a non-negative
          $A(\epsilon)$. It is possible to analytically continue similar
          anti-periodic functions, such as fermionic self-energy $\Sigma$.
          For the self-energies, it is additionally required
          that (a) the constant contribution $\Sigma(i\infty)$
          is subtracted from $\Sigma(i\omega_n)$, and (b) the zeroth spectral moment $\mathcal{N}$
          is known {\it a priori} (for an example of how $\mathcal{N}$ can be computed,
          see \cite{Potthoff1997})\footnote{An alternative approach to
          continuation of self-energies consists in constructing an auxiliary Green's function
          $G_{aux}(z) = [z + \mu - (\Sigma(z) - \Sigma(i\infty))]^{-1}$
          on the Matsubara axis, performing a \SOM run for $G_{aux}(i\omega_n)$ with $\mathcal{N}=1$ and recovering $\Sigma(\epsilon)$ from $G_{aux}(\epsilon)$ \cite{Kraberger2017}.}.

    \item Finite temperature correlation function of boson-like operators
          $B$ and $B^\dagger$,
          $\chi_B(\tau) = \langle\mathbb{T}_\tau B(\tau) B^\dagger(0)\rangle$.

          $\chi_B(\tau)$ must be $\beta$-periodic, $\chi_B(\tau+\beta) =
          \chi_B(\tau)$. Typical examples of such functions are Green's
          function of bosons $G_b(\tau) = \langle\mathbb{T}_\tau b(\tau)
          b^\dagger(0)\rangle$ and the transverse spin susceptibility
          $\chi_{+-}(\tau) = \langle\mathbb{T}_\tau S_+(\tau)S_-(0)\rangle$.

    \item Autocorrelator of a Hermitian operator. 

          The most widely used observables of this kind are
          the longitudinal spin susceptibility $\chi_{zz}(\tau) =
          \langle S_z(\tau)S_z(0)\rangle$ and the charge
          susceptibility $\chi(\tau) = \langle N(\tau)N(0)\rangle$.
          This is a special case of the previous observable kind
          with $B = B^\dagger$, and its use is in general preferred due to the
          reduced $A(\epsilon)$ definition domain (see below).

    \item Correlation function of operators at zero temperature.
          
          Strictly speaking, the imaginary time argument $\tau$ of a 
          correlation function is allowed to grow to positive infinity in the
          limit of $\beta\to\infty$.
          Therefore, one has to choose a fixed cutoff $\tau_\mathrm{max}$
          up to which the input data is available. Similarly, in the zero
          temperature limit spacing between Matsubara frequencies vanishes.
          Neither periodicity nor antiperiodicity property makes sense for observables at $\beta\to\infty$, so one can opt to use fictitious Matsubara frequencies with any of the two possible statistics and with the spacing equal to $2\pi/\tau_\mathrm{max}$.
\end{enumerate}
The spectral function $A(\epsilon)$ is defined on $(-\infty,\infty)$ for
observables (1)--(2), and on $[0,\infty)$ for observables (3)--(4).

The Matsubara representation of an imaginary time function $F(\tau)$ is assumed
to be
\begin{equation}
    F(i z_n) = \int_0^{\tilde \beta} d\tau e^{i z_n\tau} F(\tau),
\end{equation}
with $\tilde{\beta}$ being $\beta$ in the cases (1)--(3) and
$\tau_\mathrm{max}$ in the case (4). $z_n$ are fermionic Matsubara frequencies
$\omega_n = \pi(2n+1)/\beta$ in the case (1), bosonic Matsubara frequencies
$\nu_n = 2\pi n/\beta$ in the cases (2)--(3), and fictitious frequencies
$\pi(2n+1)/\tau_\mathrm{max}$ (or $2\pi n/\tau_\mathrm{max}$) in the case (4).
The Legendre polynomial representation is similar for all four observables.
\begin{equation}
    F_\ell = \sqrt{2\ell+1}\int_0^{\tilde\beta} d\tau P_\ell[x(\tau)] F(\tau),
    \quad x(\tau) = 2\tau/\tilde{\beta} - 1.
\end{equation}

Supported observables and their respective integral kernels $K(\xi,\epsilon)$ are 
summarized in Table
\ref{tab:kernels}.

\begin{table}
    \centering
    \resizebox{\columnwidth}{!}{%
    \begin{tabulary}{0.7\textwidth}{ c | c | m{2.3cm} | m{3.8cm} |}
        \cline{2-4} & \multicolumn{3}{c|}{Kernel in representation $\xi$} \\
        \hline
        \multicolumn{1}{|c|}{Observable kind} & 
        Imaginary time, $\tau$&
        Matsubara frequency, $\omega_n$/$\nu_n$&
        Legendre polynomials, $P_\ell$\\
        \hline\hline
        \multicolumn{1}{|c|}{\makecell{Green's function\\of fermions}} &
        $-\frac{e^{-\tau\epsilon}}{1+e^{-\beta\epsilon}}$ &
        $\frac{1}{i\omega_n-\epsilon}$ &
        -$\frac{\beta\sqrt{2\ell+1}(-\mathrm{sgn}(\epsilon))^\ell i_{\ell}(\beta|\epsilon|/2)}{2\cosh(\beta\epsilon/2)}$ \\
        \hline
        \multicolumn{1}{|c|}{\makecell{Correlator\\of boson-like\\operators}} &
        $\frac{1}{\pi}\frac{\epsilon e^{-\tau\epsilon}}{1-e^{-\beta\epsilon}}$ &
        $\frac{1}{\pi}\frac{-\epsilon}{i\nu_n-\epsilon}$ &
        $\frac{1}{\pi}  \frac{\beta\epsilon\sqrt{2\ell+1}(-\mathrm{sgn}(\epsilon))^\ell i_{\ell}(\beta|\epsilon|/2)}{2\sinh(\beta\epsilon/2)}$ \\
        \hline
        \multicolumn{1}{|c|}{\makecell{Autocorrelator\\of a Hermitian\\operator}} &
        $\frac{1}{\pi}
        \frac{\epsilon (e^{-\tau\epsilon}+e^{-(\beta-\tau)\epsilon})} {1-e^{-\beta\epsilon}}$ & $\frac{1}{\pi}\frac{2\epsilon^2}{\nu_n^2+\epsilon^2}$ &
        $\frac{1+(-1)^\ell}{2\pi}
        \frac{\beta\epsilon\sqrt{2\ell+1} i_{\ell}(\beta\epsilon/2)}
        {\sinh(\beta\epsilon/2)}$ \\
        \hline
        \multicolumn{1}{|c|}{\makecell{Zero temperature\\correlator}} &
        -$e^{-\tau\epsilon}$,\ $\tau\in[0;\tau_\mathrm{max}]$ &
        $\begin{array}{l}
        \frac{1}{i\pi(2n+1)/\tau_\mathrm{max}-\epsilon}\\
        \frac{1}{i2\pi n/\tau_\mathrm{max}-\epsilon}
        \end{array}$ &
        $\begin{array}{l}
        \tau_\mathrm{max}(-1)^{\ell+1}\sqrt{2\ell+1}\times\\
        i_{\ell}\left(\frac{\epsilon\tau_\mathrm{max}}{2}\right)
        \exp\left(-\frac{\epsilon\tau_\mathrm{max}}{2}\right)
        \end{array}$ \\
        \hline
    \end{tabulary}
    }
    \caption{\label{tab:kernels} List of supported integral kernels
        $K(\xi, \epsilon)$. The zero temperature correlator admits
        two frequency representations. $i_{\ell}(x)$ is the modified spherical 
        Bessel function of the first kind.}
\end{table}

\subsection{\label{sec:method}Method}

The main idea of the Stochastic Optimization Method is to use Markov chain 
sampling to minimize an objective function
\begin{equation}\label{eq:objective_function}
    D[A] = \sum_{m=1}^M |\Delta(m)|,\quad
    \Delta (m) = \frac{1}{S(m)}
    \left[
        \int\limits_{\epsilon_\mathrm{min}}^{\epsilon_\mathrm{max}}
        d\epsilon\ K(\xi_m,\epsilon) A(\epsilon) - O(\xi_m)
    \right]
\end{equation}
with respect to the spectrum $A(\epsilon)$. $\Delta(m)$ is the deviation
function for the $m$-th component of the observable. The weight factor $S(m)$
can be set to QMC error-bar estimates (if available) to stress importance of 
some input data points $O(\xi_m)$. In contrast to most other analytic
continuation methods, $S(m)$ should not, however, be identified with the
error-bars. Multiplication of all $S(m)$ by a common scalar would result in an
equivalent objective function, i.e. only the relative magnitude of different
components of $S(m)$ matters.

The integral equation (\ref{eq:fredholm}) is not guaranteed to have a unique 
solution, which means $D[A]$ has in general infinitely many minima. Most of 
these minima correspond to non-smooth spectra containing strong sawtooth noise.
SOM addresses the issue of the sawtooth instability in the following way.
At first, it generates $L$ {\it particular solutions} $\tilde A $ using a
stochastic optimization procedure. Each generated solution $\tilde A_j$ is
characterized by a value of the objective function $D[\tilde A_j]$.
A subset of relevant, or ``good'' solutions is then isolated by picking only
those $\tilde A_j$ that satisfy $D[\tilde A_j] / D_\mathrm{min}
\leq\alpha_\text{good}$. $D_\mathrm{min}$ is the absolute minimum among all
computed particular solutions, and $\alpha_\text{good}$ is a deviation
threshold factor (usually set to 2).
The final solution is constructed as a simple average of all good solutions,
and has the sawtooth noise approximately cancelled out from it.
In mathematical form, this procedure summarizes as
\begin{eqnarray}
    A(\epsilon) = \frac{1}{L_\mathrm{good}} \sum_{j=1}^L
        \theta(\alpha_\text{good} \min\{D[\tilde A_j]\} - D[\tilde A_j])
        \tilde A_j(\epsilon),\\
    L_\mathrm{good} = \sum_{j=1}^L
        \theta(\alpha_\text{good} \min\{D[\tilde A_j]\} - D[\tilde A_j]).\nonumber
\end{eqnarray}

SOM features a special way to parametrize spectra $\tilde A(\epsilon)$ as
a sum of possibly overlapping rectangles with positive weights.
\begin{align}
    \tilde A(\epsilon) &= \sum_{k=1}^K R_{\{c_k,w_k,h_k\}}(\epsilon),\\
    R_{\{c_k,w_k,h_k\}}(\epsilon) &=\label{eq:rectangle}
        h_k \theta(\epsilon - (c_k-w_k/2))\theta((c_k+w_k/2) - \epsilon).
\end{align}
$\{c_k,w_k,h_k\}$ stand for centre, width and height of the $k$-th rectangle,
respectively. The positivity constraint is enforced by requiring $w_k > 0$, 
$h_k>0$ for all $k$, while the normalization condition (\ref{eq:normalization}) 
translates into
\begin{equation}\label{eq:rect_normalization}
    \sum_{k=1}^K h_k w_k = \mathcal{N}.
\end{equation}
Sums of rectangles offer more versatility as compared to fixed energy grids
and sums of delta-peaks (as in \cite{Beach2004}). They are well suited to
represent both large scale structures, as well as narrow features of the
spectra.

The optimization procedure used to generate each particular solution
is organized as follows. In the beginning, a random configuration $\mathcal{C}_0$
(sum of $K > 0$ rectangles) is generated and height-normalized to fulfil
Eq.~(\ref{eq:rect_normalization}). Then, a fixed number $F$ of global updates
$\mathcal{C}_f\to\mathcal{C}_{f+1}$ are performed. Each global update is a 
short Markov chain of elementary updates governed by the Metropolis-Hastings 
algorithm\cite{Metropolis1953,Hastings1970} with acceptance probability
\begin{equation}
P_{t\to t+1} =
\begin{cases}
1, &D[\mathcal{C}_{t+1}] \leq D[\mathcal{C}_{t}],\\
(D[\mathcal{C}_{t}]/D[\mathcal{C}_{t+1}])^{d+1}, &D[\mathcal{C}_{t+1}] > 
D[\mathcal{C}_{t}].
\end{cases}
\end{equation}
The global update is only accepted if $D[\mathcal{C}_{f+1}] < 
D[\mathcal{C}_f]$. \SOM implements all seven types of elementary updates 
proposed by Mishchenko {\it et al} \cite{Pavarini2012}.
\begin{enumerate}
    \item Shift of rectangle;
    \item Change of width without change of weight;
    \item Change of weight of two rectangles preserving their total weight;
    \item Adding a new rectangle while reducing weight of another one;
    \item Removing a rectangle while increasing weight of another one;
    \item Splitting a rectangle;
    \item Glueing two rectangles.
\end{enumerate}
To further improve ergodicity, the Markov chain of $T$ elementary updates
is split into two stages. During the first $T_1$ elementary updates,
large fluctuations of the deviation measure are allowed by setting
$d = d_1\in (0;1]$. $d$ is then changed to $d_2\in[1;d_\text{max}]$ for the
rest of the Markov chain, so that chances to escape a local minimum are
strongly suppressed during the second stage. $T_1\in[0;T]$, $d_1$ and $d_2$
are chosen randomly and anew at the beginning of a global update.
Introduction of global updates helps accelerate convergence towards a deep
minimum of the objective function $D$, whereas MC evolution within each global update ensures possibility to escape a shallow local minimum.

An insufficient amount of global updates may pose a serious problem and result
in particular solutions $\tilde A$ that fit (\ref{eq:fredholm}) poorly.
Mishchenko suggested a relatively cheap way to check whether a given $F$ is large
enough to guarantee good fit quality \cite{Pavarini2012} (see Appendix~\ref{app:fit_quality}
for a detailed description of this recipe for real- and complex-valued
observables). Although \SOM can optionally try to automatically adjust $F$,
this feature is switched off by default. From practical calculations, we have
found that the $F$-adjustment procedure may fail to converge for input data
with a low noise level.

Another important calculation parameter is the number of particular solutions
to be accumulated. Larger $L$ lead to a smoother final solution $A(\epsilon)$
but increases computation time as $O(L)$. Setting $L$ manually to some
value from a few hundred to a few thousand range gives reasonably smooth
spectral functions in most cases. Nonetheless, \SOM can be requested to
dynamically increase $L$ until a special convergence criterion is met.
This criterion can be formulated in terms of a frequency distribution
(histogram) of the objective function $D[A]$. $L$ is considered sufficient
if the histogram is peaked near $D_\text{min} = \min\{D[\tilde A_j]\}$.
In addition to $\alpha_\text{good}$ we define $\alpha_\text{very good} <
\alpha_\text{good}$ (4/3 by default), and count those good solutions, which
also satisfy $D[\tilde A_j] / D_\text{min} < \alpha_\text{very good}$.
If nearly all good solutions (95\% or more) are very good, the histogram
is well localized in the vicinity of $D_\text{min}$, and all $\tilde A_j$
included in the final solution provide a good fit.
It is worth noting, however, that simple increase of $L$ does not in general
guarantee localization of the histogram.

\section{\label{sec:optimizations}Performance optimizations}

The most computationally intensive part of the algorithm is evaluation
of integral (\ref{eq:fredholm}) for a given configuration
$\mathcal{C} = \{c_k,w_k,h_k\}$,
\begin{align}\label{eq:sum_over_rectangles}
\int\limits_{\epsilon_\mathrm{min}}^{\epsilon_\mathrm{max}}
d\epsilon\ K(\xi_m,\epsilon) \tilde A(\epsilon) &=
\sum_{k=1}^K
\int\limits_{\epsilon_\mathrm{min}}^{\epsilon_\mathrm{max}}
d\epsilon\ K(\xi_m,\epsilon) R_{\{c_k,w_k,h_k\}}(\epsilon) =\nonumber \\ &=
\sum_{k=1}^K h_k [\Lambda(\xi_m,c_k+w_k/2) - \Lambda(\xi_m,c_k-w_k/2)],
\end{align}
where an integrated kernel $\Lambda(\xi_m,\Omega) =
\int_{\epsilon_\mathrm{min}}^\Omega K(\xi_m, \epsilon) d\epsilon$ has been introduced.
The integral over one rectangle runs from $c_k-w_k/2$ to $c_k+w_k/2$, but is computed as a difference
of two integrals: from some fixed constant ($\epsilon_{min}$) to $c_k+w_k/2$
and from the same constant to $c_k-w_k/2$. The lower integration limit in the definition of
$\Lambda(\xi_m,\Omega)$ is irrelevant and can be set to an arbitrary number, e.g. 0.
This has the benefit that only one integral needs to be known/evaluated for all possible upper limits.

In the Matsubara representation, the integrated kernels $\Lambda(\xi_m,\Omega)$ are simple
analytic functions (see Appendix \ref{app:int_kernels_imfreq}). For the
imaginary-time and Legendre representations, however, there are no closed form
expressions for $\Lambda(\xi_m,\Omega)$. Doing the integrals with a
numerical quadrature method for each rectangle and upon each proposed elementary
update would be prohibitively slow. \SOM boasts two major optimizations that
strongly alleviate the problem.

\begin{figure}
\centering
\includegraphics[scale=0.7]{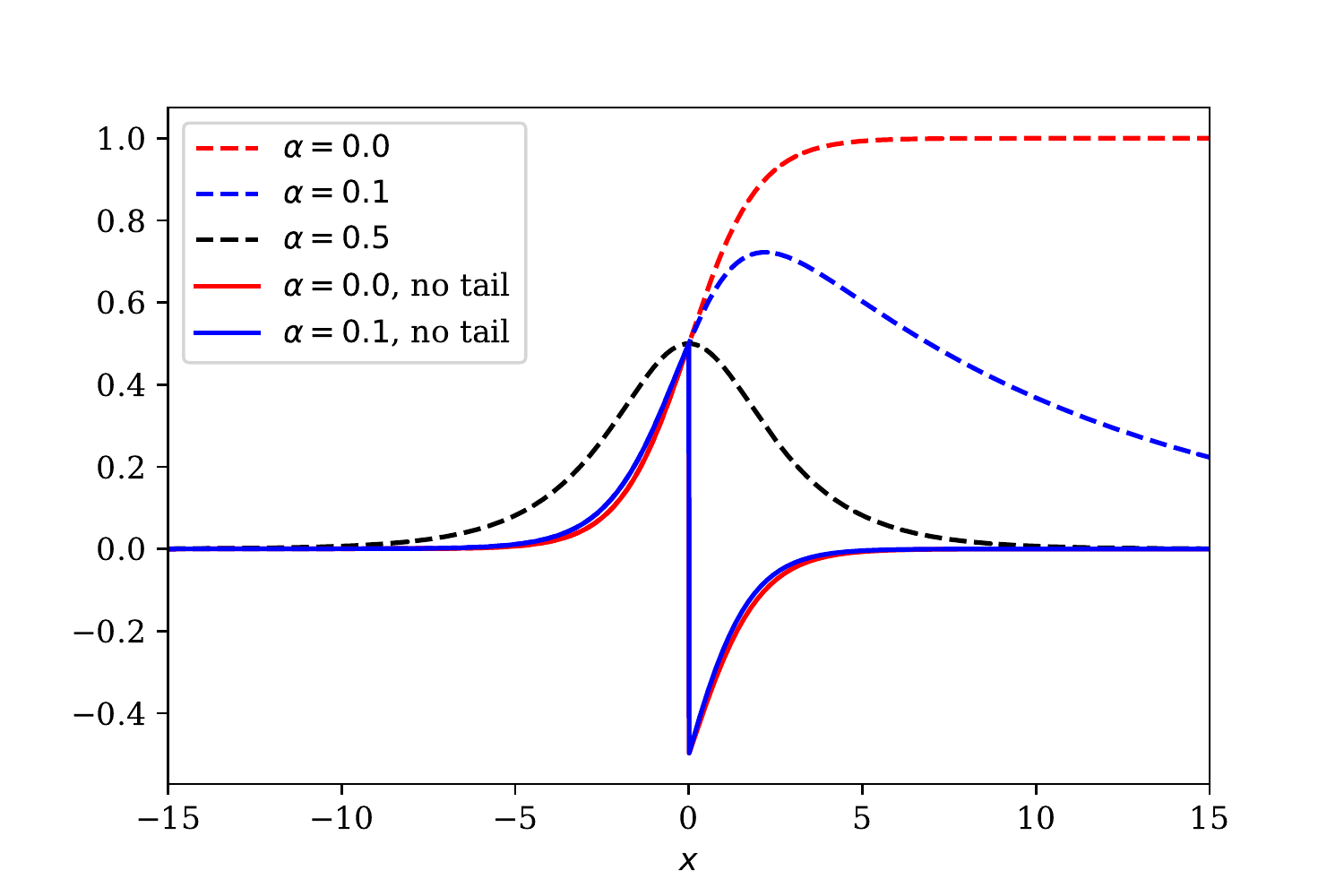}
\caption{\label{fig:imtime_lambda}Illustration of $\Lambda(\tau,\Omega)$
    evaluation problem. Dashed lines: Integrand from (\ref{eq:imtime_lambda})
    for different values of $\alpha=\tau/\beta$. Solid lines: The same 
    integrand with the tail contribution $t_\alpha(x)$ subtracted.
}
\end{figure}

At first, it uses pre-computation and special interpolation schemes to quickly
evaluate $\Lambda(\xi_m,\Omega)$. As an example, let us consider the kernel for
the fermionic GF in the imaginary time (for definition of kernel $K(\tau,\epsilon)$
see the upper left cell of Table \ref{tab:kernels}),
\begin{equation}\label{eq:imtime_lambda}
\Lambda(\tau;\Omega) = -\frac{1}{\beta}
\int\limits_0^{\beta\Omega}\frac{e^{-(\tau/\beta)x}}{1+e^{-x}}dx
\end{equation}
($\epsilon$ has been substituted by $x/\beta$, the lower integration limit is set to 0).
The integral on the right hand side is analytically doable only for
a few values of $\alpha=\tau/\beta$, namely $\alpha=0,1/2,1$.
For all other $\alpha\in(0;1/2)\cup(1/2;1)$ it has to be done numerically and 
approximated using a cubic spline interpolation. The spline interpolation
would be easy to construct if the integrand were localized near $x=0$ for
all $\alpha$. Unfortunately, this is not the case. The integrand develops
a long ``tail'' on the positive/negative half-axis as $\alpha$ approaches
0/1 respectively. The length of this tail scales as $\alpha^{-1}$ (or 
$(1-\alpha)^{-1}$), which would require an excessively large number of
spline knots needed to construct a reliable approximation.
This issue is solved by subtracting an exponential tail contribution 
$t_\alpha(x)$ from the integrand, such that the difference is well localized, 
and an integral of $t_\alpha(x)$ is trivial (Fig.~\ref{fig:imtime_lambda}).
\begin{equation}
\Lambda(\tau;\Omega) = -\frac{1}{\beta} \left[
\theta(-\Omega)S^-_\alpha(\beta\Omega) +
\theta(\Omega)S^+_\alpha(\beta\Omega) +
T_\alpha(\beta\Omega)
\right],
\end{equation}
\begin{equation}
S^-_\alpha(z) = -\int\limits_z^0
\left[\frac{e^{-\alpha x}}{1+e^{-x}} - t_\alpha(x)
\right] dx,\quad
S^+_\alpha(z) = \int\limits_0^z
\left[\frac{e^{-\alpha x}}{1+e^{-x}} - t_\alpha(x)
\right] dx,
\end{equation}
\begin{equation}
T_\alpha(z) = \int\limits_0^z t_\alpha(x) dx.
\end{equation}

For each $\alpha_m = \tau_m/\beta$, the integrals $S^-_\alpha(z)$ and
$S^+_\alpha(z)$ are expressed in terms of special functions or rapidly
convergent series and are precomputed on a uniform grid with a fixed number of
$z$-points. The precomputed values provide data to construct the cubic spline
interpolation between those points, and $T_\alpha(z)$ are simple analytical
functions that are quick to evaluate. The length of the segment, on which the
splines are defined, is chosen such that functions $S^\pm_\alpha(z)$ can be
considered constant outside it. Appendix \ref{app:int_kernels_imtime}
contains relevant expressions and necessary implementation details for all
observable kinds.

A somewhat similar approach is taken to evaluate integrated Legendre kernels.
Once again, let us take the fermionic GF as an example
(upper right cell of Table \ref{tab:kernels}),
\begin{equation}
\Lambda(\ell;\Omega) =
-\int\limits_0^{\Omega}\frac{\beta\sqrt{2\ell+1}(-\mathrm{sgn}(\epsilon))^\ell
i_{\ell}\left(\frac{\beta|\epsilon|}{2}\right)}{2\cosh(\beta\epsilon/2)} d\epsilon =
(-\mathrm{sgn}(\Omega))^{\ell+1}\sqrt{2\ell+1}
\int\limits_0^{|\Omega|\beta/2} \frac{i_\ell(x)}{\cosh(x)} dx,
\end{equation}
where $i_\ell(x)$ is the modified spherical Bessel function of the first kind
\cite{spherical_bessel}.
The integrand $\frac{i_\ell(x)}{\cosh(x)}$ is rather inconvenient
for numerical evaluation, because both the Bessel function and the
hyperbolic cosine grow exponentially. Moreover, the integral itself grows
logarithmically for $|\Omega|\beta\to\infty$, which makes naive pre-computation
and construction of a spline infeasible. This time, our evaluation scheme is
based on the following expansion of the integrand,
\begin{equation}\label{eq:il_cosh_series}
\frac{i_\ell(x)}{\cosh(x)} =
\frac{e^x}{e^x+e^{-x}}\sum_{n=0}^\ell(-1)^n
\frac{a_n(\ell+1/2)}{x^{n+1}} +
\frac{e^{-x}}{e^x+e^{-x}}\sum_{n=0}^\ell(-1)^{\ell+1}
\frac{a_n(\ell+1/2)}{x^{n+1}},
\end{equation}
with $a_n(\ell+1/2)$ being coefficients of the Bessel polynomials
\cite{carlitz1957},
\begin{equation}
a_n(\ell+1/2) = \frac{(\ell+n)!}{2^n n!(\ell-n)!}.
\end{equation}
For large $x$ (high-energy region), the integrand can be approximated as
\begin{equation}\label{eq:il_cosh_series_high}
\frac{i_\ell(x)}{\cosh(x)} \approx
\sum_{n=0}^\ell(-1)^n \frac{a_n(\ell+1/2)}{x^{n+1}}.
\end{equation}
In the low-energy region we do the integral $F^<(z) \equiv \int_0^z
\frac{i_\ell(x)}{\cosh(x)} dx$ on a fixed $z$-mesh, $z\in[0;x_0]$, using the
adaptive Simpson's method. Results of the integration are used to construct a
cubic spline interpolation of $F^<(z)$. For each $\ell$ the boundary value
$x_0$ between the low- and high-energy regions is found by comparing values
of (\ref{eq:il_cosh_series}) and (\ref{eq:il_cosh_series_high}).
In the high-energy region we use the asymptotic form
(\ref{eq:il_cosh_series_high}) to do the integral,
\begin{equation}
F^>(z)|_{z>x_0} = F^<(x_0) +
\left.\left\{
\log(x) +
\sum_{n=1}^\ell (-1)^{n+1}\frac{a_n(\ell+1/2)}{x^n n}
\right\}\right|_{x_0}^z.
\end{equation}
Implementation details for other Legendre kernels can be found in
Appendix~\ref{app:int_kernels_legendre}.

The second optimization implemented in \SOM consists in aggressive caching of
rectangles'
contributions to the integral (\ref{eq:fredholm}). As it is seen
from (\ref{eq:sum_over_rectangles}), the integral is a simple sum over all
rectangles in a configuration. Thanks to the fact that elementary updates
affect at most two rectangles at once, it makes sense to store values of every
individual term in the sum. In the proposal phase of an update, evaluation of
the integrated kernel is then required only for the affected rectangles. If the
update is accepted, contributions of the added/changed rectangles are stored in
a special cache and can be reused in a later update. Since the size of a configuration
does not typically exceed 100, and $M\leq1000$, the cache requires only a
moderate amount of memory.

Besides the two aforementioned optimizations, \SOM can take advantage of
MPI parallel execution. Generation of particular solutions is ``embarrassingly
parallel'', because every solution is calculated in a completely independent
manner from the others. When a \SOM calculation is run in the MPI context,
the accumulation of particular solutions is dynamically distributed among
available MPI ranks.

\section{\label{sec:usage}Usage}

\subsection{\label{sec:example}Basic usage example}

Running \SOM to analytically continue input data requires writing a simple
Python script. This usage method is standard for \TRIQS applications.
We refer the reader to Sec. 9.3 of \cite{TRIQS} for instructions on how to 
execute Python scripts in the \TRIQS environment. Details of the script will 
vary depending on the physical observable to be continued, and its 
representation. Nonetheless, a typical script will have the following basic 
parts.

\begin{itemize}
\item Import \TRIQS and \SOM Python modules.
\begin{pylisting}
# Green's function containers used to store input and output data.
from pytriqs.gf.local import *

# HDFArchive interface to HDF5 files.
from pytriqs.archive.hdf_archive import HDFArchive, HDFArchiveInert

# HDF5 archives must be modified only by one MPI rank.
# is_master_node() checks we are on rank 0.
from pytriqs.utility.mpi import is_master_node

# bcast() broadcasts its argument from the master node to all others
from pytriqs.utility.mpi import bcast

# Main SOM class.
from pytriqs.applications.analytical_continuation.som import Som
\end{pylisting}

\item Load the observable to be continued from an HDF5 archive.
\begin{pylisting}
# On master node, open an HDF5 file in read-only mode.
arch = HDFArchive('input.h5', 'r') if is_master_node() else HDFArchiveInert()

# Read input Green's function and broadcast it to all nodes.
# arch['g'] must be an object of type GfImTime, GfImFreq or GfLegendre.
g = bcast(arch['g'])
\end{pylisting}
This step can be omitted or modified if the input data comes from a different
source. For instance, it could be loaded from text files or generated in the
very same script by a quantum Monte-Carlo impurity solver, such as \TRIQS/\CTHYB
\cite{CTHYB}. Only the values stored in the \py{g.data} array will be used by
\SOM, while any high frequency expansion information (\py{g.tail}) will be
disregarded. If \py{g} is matrix-valued (or, in \TRIQS's terminology, has a
target shape bigger than 1x1), \SOM will only construct analytic continuation
of the diagonal matrix elements.

\item Set the importance function $S(m)$
\begin{pylisting}
# Create a container for 'S' by copying 'g'.
S = g.copy()
# Set all elements of 'S' to a constant.
S.data[:] = 1.0
\end{pylisting}

In this example we assume that all elements of $g(m)$ are equally important.
Alternatively, one could read the importance function from an HDF5 archive or
another source.

\item Construct \py{Som} object.
\begin{pylisting}
# Create Som object
cont = Som(g, S, kind = "FermionGf", norms = 1.0)
\end{pylisting}

The argument \py{kind} must be used to specify the kind of the physical observable 
in question. Its recognized values are \py{FermionGf} (Green's function of 
fermions), \py{BosonCorr} (correlator of boson-like operators), 
\py{BosonAutoCorr} (autocorrelator of a Hermitian operator),
\py{ZeroTemp} (zero temperature correlator). The optional argument \py{norms}
is a list containing expected norms $\mathcal{N}$ (Eq.~\ref{eq:normalization})
of the spectral functions to be found, one positive real number per one
diagonal element of \py{g}.
Instead of setting all elements of \py{norms} to the same constant \py{x}, one 
may simply pass \py{norms = x}. By default, all norms are set to 1, which is 
correct for the fermionic Green's functions. However, adjustments would 
normally be needed for self-energies and bosonic correlators/autocorrelators.

\item Set simulation parameters.
\begin{pylisting}
# Dictionary containing all simulation parameters
params = {}

# SOM will construct a spectral function
# within this energy window (mandatory)
params['energy_window'] = (-5.0,5.0)

# Number of particular solutions to accumulate (mandatory).
params['l'] = 1000

# Set verbosity level to the maximum on MPI rank 0,
# and silence all other ranks
params['verbosity'] = 3 if is_master_node() else 0

# Do not auto-adjust the number of particular solutions to accumulate
# (default and recommended).
params['adjust_l'] = False

# Do not auto-adjust the number of global updates (default and recommended).
params['adjust_f'] = False

# Number of global updates per particular solution.
params['f'] = 200

# Number of local updates per global update.
params['t'] = 500

# Accumulate histogram of the objective function values,
# useful to analyse quality of solutions.
params['make_histograms'] = True
\end{pylisting}

In Section \ref{sec:parameters} we provide a table of main accepted
simulation parameters.

\item Run simulation.
\begin{pylisting}
cont.run(**params)
\end{pylisting}

This function call is normally the most expensive part of the script.

\item Extract solution and reconstructed input.
\begin{pylisting}
# Evaluate the solution on a real frequency mesh.
g_w = GfReFreq(window = (-5.0, 5.0), n_points = 1000, indices = g.indices)
g_w << cont

# Imaginary time/frequency/Legendre data reconstructed from the solution.
g_rec = g.copy()
g_rec << cont
\end{pylisting}
\py{g_w} is the retarded fermionic Green's function of the real frequency 
corresponding to the input \py{g}. Its imaginary part is set to $-\pi 
A(\epsilon)$, whereas the real part is computed semi-analytically using the 
Kramers--Kronig relations (there are closed form expressions for a contribution
of one rectangle to the real part). The relation between \py{g_w} and
$A(\epsilon)$ slightly varies with the observable type; relevant details are to
be found in the online documentation.

The high frequency expansion coefficients will be computed from the 
sum-of-rectangles representation of $A(\epsilon)$ and written into \py{g_w} as 
well.
If \py{g_w} is constructed with a wider energy window compared to that from
\py{params}, $A(\epsilon)$ is assumed to be zero at all ``excessive'' energy
points.

The reconstructed function \py{g_rec} is the result of the substitution of
$A(\epsilon)$ back into the integral equation (\ref{eq:fredholm}).
The correctness of results should always be controlled by comparing \py{g} with 
\py{g_rec}.

\item Save results to an HDF5 archive.
\begin{pylisting}
# On master node, save results to an archive
if mpi.is_master_node():
    with HDFArchive("results.h5",'w') as ar:
        ar['g'] = g
        ar['g_w'] = g_w
        ar['g_rec'] = g_rec
        # Save histograms for further analysis
        ar['histograms'] = cont.histograms
\end{pylisting}

\end{itemize}

The output archive can be read by other \TRIQS scripts and outside utilities,
in order to analyse, post-process and visualize the resulting spectra.
More elaborate examples of \SOM Python scripts can be found on the official
documentation page, \url{http://krivenko.github.io/som/documentation.html}{}.

\subsection{\label{sec:parameters}Simulation parameters}

Table (\ref{tab:main_parameters}) contains main simulation parameters 
understood by \py{run()} method of class \py{Som}. More advanced parameters
intended for experienced users are listed in Appendix \ref{app:parameters}.

\begin{table}[h!]
\centering
\resizebox{\columnwidth}{!}{%
\begin{tabularx}{\textwidth}{|c|c|c|X|}
    \hline
    Parameter Name & Python type & Default value & \makecell{Meaning} \\
    \hline\hline
    \py{energy_window} & \py{(float,float)} & N/A &
    Lower and upper bounds of the energy window.
    Negative values of the lower bound will be reset to 0 in
    \py{BosonAutoCorr} and \py{ZeroTemp} modes.\\
    \hline
    \py{max_time} & \py{int} & \py{-1} (unlimited) &
    Maximum \py{run()} execution time in seconds.\\
    \hline
    \py{verbosity} & \py{int} & \makecell{\py{2} on MPI rank 0,\\
    \py{0} otherwise} &
    Verbosity level (maximum level --- 3).\\
    \hline
    \py{t} & \py{int} & \py{50} & Number of elementary updates per global
    update ($T$). Bigger values may improve ergodicity.\\
    \hline
    \py{f} & \py{int} & \py{100} & Number of global updates ($F$).
    Bigger values may improve ergodicity. Ignored if \py{adjust_f=True}.\\
    \hline
    \py{adjust_f} & \py{bool} & \py{False} &
    Adjust the number of global updates automatically.\\
    \hline
    \py{l} & \py{int} & \py{2000} &
    Number of particular solutions to accumulate ($L$). Bigger values reduce
    the sawtooth noise in the final solution. Ignored if \py{adjust_l=True}.\\
    \hline
    \py{adjust_l} & \py{bool} & \py{False} &
    Adjust the number of accumulated particular solutions automatically.\\
    \hline
    \py{make_histograms} & \py{bool} & \py{False} &
    Accumulate histograms of objective function values.\\
    \hline
\end{tabularx}
}
\caption{\label{tab:main_parameters}Main \py{run()} parameters.}
\end{table}

\clearpage
\section{\label{sec:results}Illustrational results}

\subsection{\label{sec:results_intro}Testing methodology}

We apply \SOM to a few synthetic test cases where the exact spectral function is
known. These tests serve as an illustration of \SOM's capabilities and as 
guidance to the prospective users. Being based on a Markov chain algorithm with 
a fair number of adjustable parameters, \SOM may suffer from ergodicity 
problems and produce characteristic spectral artefacts if the parameters are 
not properly chosen.
It is therefore instructive to consider analytic continuation problems
that can potentially pose difficulty to \SOM, and study the effect of the various
input parameters on the quality of the solution. Throughout this section,
we use arbitrary units for all energies and frequencies ($\hbar=1$), while
imaginary time $\tau$ and inverse temperature $\beta$ are measured in $[a.u.^{-1}]$.

\begin{figure}
\centering
\includegraphics{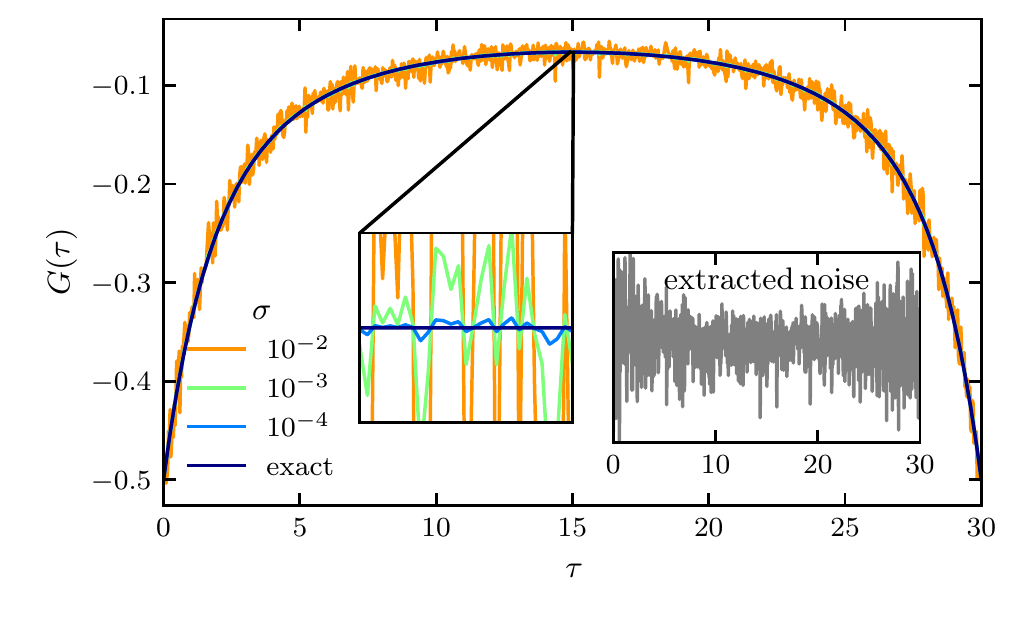}
\caption{\label{fig:noise} Extraction of noise from the imaginary time 
Green's function $G(\tau)$ computed by \TRIQS/\CTHYB QMC solver, and its 
subsequent rescaling.
}
\end{figure}

To perform most of the tests, we have devised a special procedure allowing
to model stochastic noise characteristic to fermionic Green's functions
$G(\tau)$ measured by QMC solvers. The procedure involves running a single-loop
DMFT\cite{DMFTReview} calculation for a non-interacting single-band
tight-binding model on a Bethe lattice using the \TRIQS/\CTHYB solver. The exact
Green's function of this model $G_\text{Bethe}^\text{ex}(\tau)$ is easily
computed from the
well known semi-elliptic spectral function. The impurity solver
will introduce some noise $\tilde\eta(\tau)$ that can easily be 
estimated from the accumulation result $G_\text{Bethe}^\text{MC}(\tau)$ as
\begin{equation}
\label{eq:noise}
\tilde\eta(\tau) =
G_\text{Bethe}^\text{MC}(\tau) - G_\text{Bethe}^\text{ex}(\tau),
\end{equation}
and normalized according to
\begin{equation}
\eta(\tau) = \tilde\eta(\tau) \left/
\sqrt{\frac{1}{M}\sum_m \left(\tilde\eta(\tau_m) -
\frac{1}{M}\sum_{m'}\tilde\eta(\tau_{m'})
\right)^2}
\right..
\end{equation}

We ensure that values of the noise are uncorrelated between different time 
slices $\tau_m$ by taking a sufficiently large number of Monte Carlo updates 
per measurement. Eventually, the extracted and normalized noise is rescaled and
added to a model Green's function to be continued with \SOM 
(Fig.~\ref{fig:noise}),
\begin{equation}
\label{eq:gnoise}
G^\text{noisy}_\sigma(\tau) = G^\text{ex}(\tau) + \sigma \eta(\tau).
\end{equation}
By using the described procedure, we hope to reproduce a more
realistic noise distribution and its dependence on $\tau$ observed in QMC runs,
as compared to a synthetically generated Gaussian noise with a fixed dispersion.

In the case of a susceptibility (two-pole model, see below), we simply add
Gaussian noise to $\chi(i\nu_m)$. It is independently generated for each 
$\nu_m$ 
with zero mean and equal dispersion $\sigma$.

\subsection{\label{sec:results_models}Comparison of model spectra}

\begin{figure}
    \centering
    \includegraphics{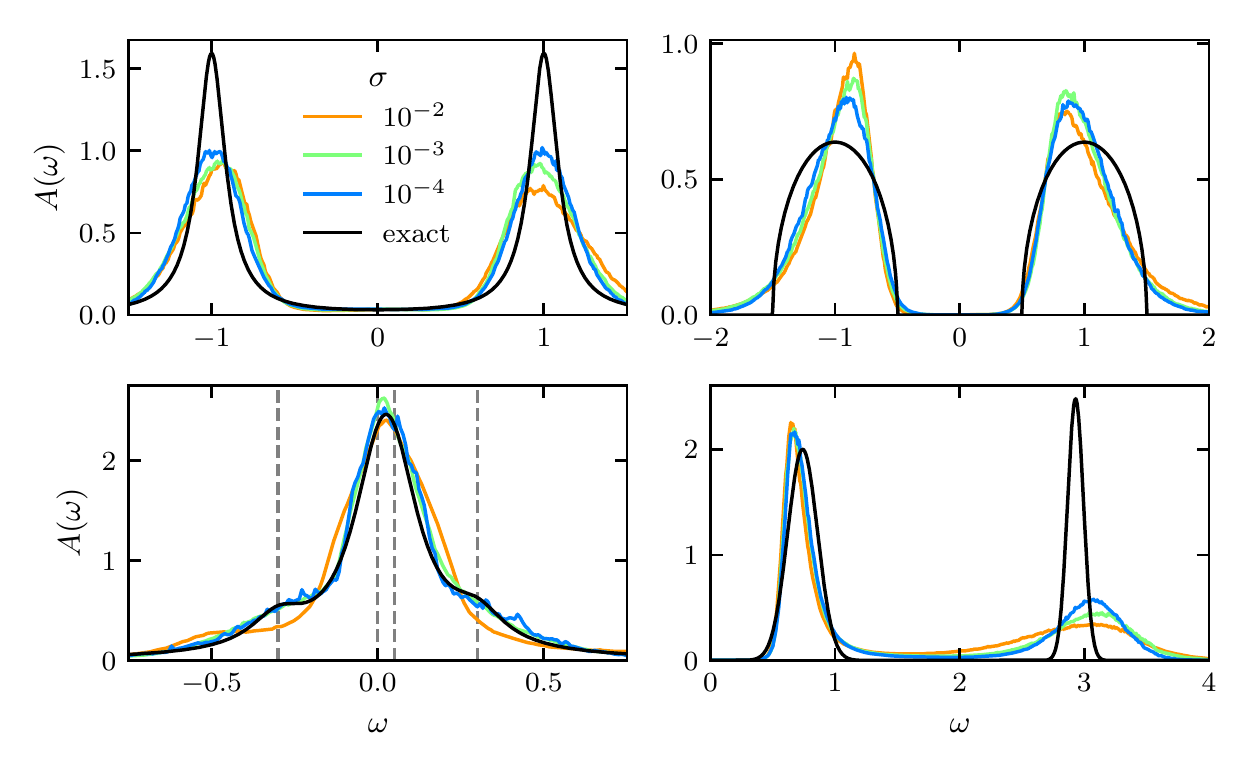}
    \caption{\label{fig:sigma_convergence} Spectral functions
        produced by \SOM with $F = 1500$ depending on the input noise level $\sigma$.
        The corresponding models are the Hubbard atom (top left), the 
        double Bethe 
        lattice (top right), an asymmetric metal (bottom left), and the 
        zero-temperature Fermi polaron (bottom right). The frequency positions of the asymmetric metal's superposed Lorentzians are marked by dashed lines.
}
\end{figure}
Results of tests with four model spectra and various levels of added noise are 
shown in Fig.~\ref{fig:sigma_convergence}.
In all tests presented in that figure, the inverse 
temperature has been set to $\beta = 30$ ($\tau_{max} = 30$ for the 
zero-temperature kernel).
All curves have been calculated from the imaginary time input data on a grid 
with $500$ $\tau$-points ($G^\text{ex}(\tau)$ are computed from analytically 
known spectral functions via (\ref{eq:fredholm})).
The importance function has been chosen to be a constant in all simulations. 
The number of global updates is set to $F=1500$ and the number of elementary 
updates is $T = 250$. The remaining parameters are set to \SOM defaults, if not 
stated otherwise.

The model studied first is the single Hubbard atom with the Coulomb repulsion
strength $U=2$ at half-filling. Its spectral function comprises two
discrete levels at $\pm U/2$ corresponding to excitations through empty and 
doubly occupied states. For a general complex frequency $z$, the Green's 
function of the Hubbard atom reads
\begin{gather}
  \label{eq:hubbardatom}
  G(z) = \frac{1/2}{z+U/2+i\delta} + \frac{1/2}{z-U/2+i\delta},
\end{gather}
where $\delta=0.1$ is the resonance broadening term. The analytic
continuation with \SOM shows that strong noise smears the calculated spectral
function, the peak height is diminished and the spectral weight is spread
asymmetrically around the peak. The smearing at large frequencies is more
pronounced. The peak position converges correctly to the exact solution as
$\sigma$ decreases.

The second model is the non-interacting double Bethe lattice \cite{Moeller1999},
\begin{gather}
  \label{eq:doublebethe}
  G(z) = \zeta(z-t_\perp) + \zeta(z+t_\perp), \\
  \zeta(z) = \frac{z - i\sign\Im(z) \sqrt{4 t_b^2 - z^2}}{2 t_b^2}.
\end{gather}
The specific feature of this model is a gapped symmetric spectrum with sharp
band edges. Splitting between the semi-elliptic bands is given by $|2t_\perp| =
2$, and their bandwidths are $4t_b = 1$.
Top right part of Fig.~\ref{fig:sigma_convergence} shows that at large $\sigma$
the particle--hole asymmetry introduced by the noise can also break
the symmetry of the solution. As the noise level goes down, the \SOM 
solution approaches the exact one, but the shape of the bands remains somewhat
Lorentzian-like, which leads to an underestimated gap.

As an example of a gapless spectrum (Fig.~\ref{fig:sigma_convergence}, bottom
left), we take a Green's function that produces a sum of four Lorentzian peaks, i.e.
\begin{equation}
  \label{eq:asymmetricmetal}
  G(z) = \sum_i \frac{c_i}{z-b_i+i\delta}
\end{equation}
with $c = \left(0.1,0.4,0.4,0.1\right)$, $b=\left(-0.3,0,0.05,0.3\right)$ and
$\delta=0.1$.
The Lorentzians overlap and form a single peak that has two shoulders. The peak
doesn't have a mirror symmetry with respect to the vertical axis and its centre
is not at the Fermi level. The shoulders are not resolved for the largest noise
level of $\sigma = 10^{-2}$, but for smaller noise levels $\sigma \in \{10^{-3},
10^{-4}\}$ both shoulders are correctly detected by \SOM. The spectra with 
$\sigma \in \{10^{-3}, 10^{-4}\}$ look notably more noisy.
The origin of this effect is \SOM's ability to find the best particular 
solution that is a very good fit to the input data (with a very small 
$D_\text{min}$). With a fixed $D[A]/D_\text{min}$ threshold, only a few 
particular solutions are considered for inclusion in the final solution. Hence, 
the less smooth curves.

The fourth model (Fig.~\ref{fig:sigma_convergence}, bottom right) appears in 
the context of the Fermi polaron problem, and it has previously been used in
\cite{Goulko2017} to assess performance of other continuation methods. It 
consists of two peaks modelled via Gaussians,
\begin{equation}
  \label{eq:fermipolaron}
  G(z) = \frac{1}{N} \sum_i c_i e^{-\frac{(z-a_i)^2}{b_i}}
\end{equation}
with $c = \left(0.62, 0.41\right)$, $a = \left(0.74, 2.93\right)$ and $b =
\left(0.12, 0.064\right)$. $N$ is used to set the spectral norm to $1$. Here,
we apply the zero-temperature imaginary time kernel with $\tau_{max} = 30$. This case is difficult as it contains sharp features over
a broad energy range. We observe that the low-energy peak can be relatively
well resolved. However, the high-energy peak is strongly broadened with the 
width dependent of the noise level. Positions of both peaks are reproduced with 
a reasonable accuracy of $\sim 5 \%$.

\subsection{\label{sec:results_f}Effect of $F$}

\begin{figure}
\centering
\includegraphics{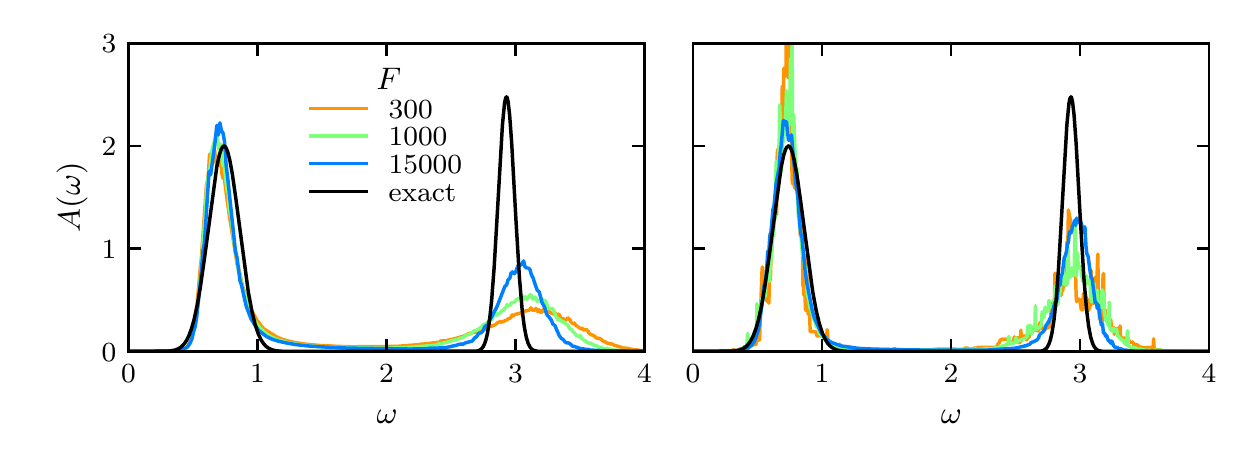}
\caption{\label{fig:ferm_f_tau_leg} Effect of the number of global updates $F$
 on the spectral functions produced by \SOM.
Shown are spectral functions for the Fermi polaron model obtained from the imaginary time (left)
and Legendre (right) input data.}
\end{figure}

We have observed in practical calculations that the choice of the amount of global 
updates $F$ can drastically affect ergodicity of \SOM Markov chains and, 
eventually, the output spectra. 
Fig.~\ref{fig:ferm_f_tau_leg} shows evolution of \SOM spectral functions with 
$F$ for the Fermi polaron model. Parameters other than $F$ are kept at their 
default values, i.e. $T=50$, $L=2000$, and the noise level is set to $\sigma = 
10^{-4}$.
The Fermi polaron spectra obtained from the imaginary time representation
(Fig.~\ref{fig:ferm_f_tau_leg}, left) demonstrates that \SOM has generally
no problem resolving the low-energy peak regardless of $F$. The high-energy
peak, however, significantly changes its shape and position as $F$ grows,
slowly approaching the reference curve.

The situation is rather different, when the Legendre representation is used
(Fig.~\ref{fig:ferm_f_tau_leg}, right). As before, the noise is created and 
added to the Green's function in the imaginary time representation. 
Subsequently, it is transformed to the Legendre basis and passed to \SOM. The 
number of Legendre coefficients used here is $N_\ell=50$. Positions of both 
peaks do not change much with $F$. For sufficiently large $F$,
their width and height are visibly closer to the exact solution as compared to
the imaginary time results. A likely cause of this difference is the truncation
of the Legendre expansion and the corresponding noise filtering.
In contrast, for small $F$ the peaks contain relatively strong noise. This 
effect has been found with the asymmetric metal model for large $\sigma$, too. 
It occurs
when the final solution is dominated by a few particular solutions with very
small $D$, i.e. when Markov chains suffer from poor ergodicity and struggle
to escape an isolated deep minimum of the objective function.

\begin{figure}
\centering
\includegraphics{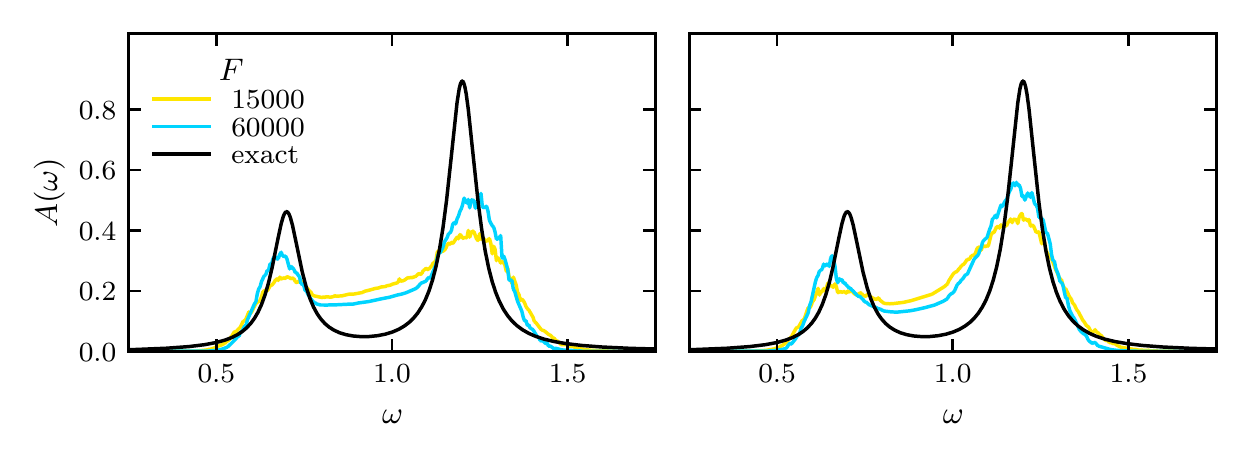}
\caption{\label{fig:twop_f_iw_leg} Effect of the number of global updates $F$ on the spectral functions produced by \SOM.
Shown are spectral functions for the two-pole model obtained from the Matsubara 
frequency (left)
and Legendre (right) input data.}
\end{figure}

In Fig.~\ref{fig:twop_f_iw_leg},
a similar comparison is presented for the two-pole susceptibility model\cite{Schott2016},
\begin{gather}
  \label{eq:twopolemodel}
  \chi(z) = \frac{1}{\chi_0}\sum_i \frac{c_i}{z^2 - b^2_i},\\
  \chi_0 = -\sum_i \frac{c_i}{b^2_i},\nonumber
\end{gather}
with $c=\left(0.1, 0.335663\right)$, $b=\left(0.7, 1.2\right)$. We use the
Hermitian autocorrelator kernel for Matsubara frequencies and add uncorrelated
Gaussian noise with the standard deviation $\sigma = 10^{-4}$. Other parameters
are $\beta = 50$, $T=50$ and $L=2000$. For the same number of global
updates, we obtain a double-peak structure (Fig.~\ref{fig:twop_f_iw_leg}, left)
similar to that shown in Fig.~2 of \cite{Schott2016}.
Additionally, we transform the same set of noisy
input data into the Legendre representation ($N_\ell=50$) and apply the corresponding
kernel (Fig.~\ref{fig:twop_f_iw_leg}, right). As in the case of the Fermi polaron, use of the
Legendre basis results in slightly more pronounced peaks. However, their positions are reproduced
slightly better by the Matsubara kernel.

\subsection{\label{sec:results_diagnostics}Diagnostics}

\begin{figure}
\centering
\includegraphics{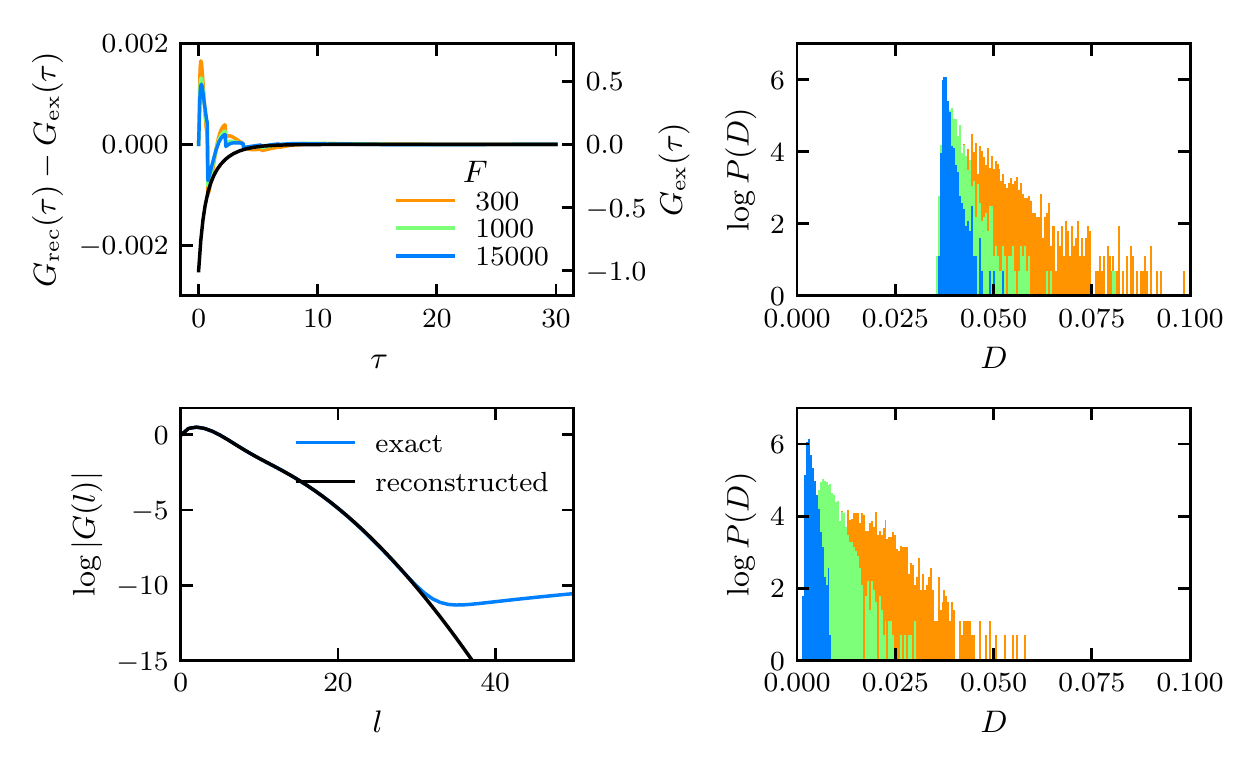}
\caption{\label{fig:hist_rec} Top, left: Colourful lines/left axis show the
difference of the \SOM-reconstructed Green's function and the exact Green's
function of the Fermi polaron model, continued using the imaginary time
kernel. Black curve/right axis shows the exact Green's function itself.
Bottom, left: The reconstructed Green's function and the exact Green's function 
in the Legendre basis (same model, $F=15000$).
Right: Corresponding histograms of objective function values.}
\end{figure}

After a solver run, \SOM does not only provide the continued function, but also 
the reconstructed input function as well as a histogram of the objective 
function. These features are mainly for debugging purposes and for an estimation 
of result's quality.
It is obvious that the reconstructed Green's function should ideally be a smooth version of the input Green's function. 
In the upper left part of
Fig.~\ref{fig:hist_rec} we plot the difference between the exact Green's
function of the Fermi polaron model and its \SOM reconstruction after an
imaginary time simulation (the corresponding spectrum is shown in 
Fig.~\ref{fig:ferm_f_tau_leg}). First of all, we observe that the difference 
shrinks with increasing $F$. Furthermore, the main difference comes from the 
small imaginary times. The Green's function has its
largest values there, and so does the noise. At larger $\tau$-points it has
less pronounced features or is even featureless. However, $\tau_\text{max}$ has to be set
sufficiently large, i.e. the result should converge and be independent of its actual value.

The Legendre basis has been identified as a useful tool to filter noise by 
cutting off large expansion coefficients.\cite{Boehnke2011}
In Fig.~\ref{fig:hist_rec} (bottom left) we provide a similar comparison of
exact and reconstructed Green's functions in the Legendre representation. They
lie on top of each other everywhere except for large-order Legendre
coefficients. This kind of plots may give a hint at how many Legendre
polynomial coefficients should be retained to filter out the noise without losing
relevant information in the input data. 

The objective function histogram helps determine whether the choice of $F$
and $T$ is adequate. The histogram is usually stretched when $T$ and/or
$F$ are too small, and should converge to a sharper peak shape for larger
values. Objective function histograms for both
used kernels are plotted in the right part of Fig.~\ref{fig:hist_rec}.
It is clearly seen that increased $F$ lead to localization of the histograms
and, as a result, to more efficient accumulation of particular solutions.

\subsection{\label{sec:results_remarks}Final remarks}

In conclusion, we would like to make some remarks about the rest of the main
simulation parameters. The energy window must be chosen such that the entire
spectral weight fits into it, but within the same order of magnitude with the
bandwidth. Extremely wide energy windows may be wasteful as they are likely to
cause low acceptance rates in the Markov chains. The effect of increased $L$
has not been explicitly studied here, but it is similar to that of a moving
average noise filter. It is, however, not recommended to post-process (smoothen)
\SOM-generated spectra using such a filter, because it can cause artefacts
that also depend on the moving frame size.
Regarding efficiency, it is worth noting that simulations with the Legendre 
kernels typically require less CPU time, mainly because of the different dimensions of the input 
data arrays (a few dozen Legendre coefficients versus $10^2-10^3$ 
$\tau$-points/Matsubara frequencies).
To give the reader a general idea of how expensive
\SOM simulations are, we roughly estimated the CPU time required
to produce some plots in this section. It took 136 core-hours to 
obtain one spectral curve for the Hubbard atom, the double Bethe 
lattice, and the asymmetric metal (Fig.~\ref{fig:sigma_convergence}).
The Fermi polaron simulations were a lot cheaper, 19 core-hours.
The most expensive curve ($F=15000$) in Fig.~\ref{fig:ferm_f_tau_leg}
took 20 core-hours with the imaginary time kernel, and 8 core-hours with the
Legendre kernel. Similarly, the $F=60000$ curve of Fig.~\ref{fig:twop_f_iw_leg}
required $25$ and $14$ core-hours with the imaginary time and the Legendre
kernels respectively. The required run time scales nearly linearly with $F$.

\section{\label{sec:starting}Getting started}

\subsection{\label{sec:obtaining}Obtaining \SOM}

The \SOM source code is available publicly and can be obtained from a GitHub
repository located at \url{https://github.com/krivenko/som}. It is recommended
to always use the `master' branch of the repository, as it contains the most recent bug fixes and new features being added to \SOM.

\subsection{\label{sec:installation}Installation}

The current version of \SOM requires the \TRIQS library 1.4.2 as a prerequisite.
An installation procedure of \TRIQS is outlined in Sec. 9.2 of
\cite{TRIQS}. More detailed step-by-step instructions are available on the
\TRIQS documentation website
(\url{https://triqs.github.io/triqs/1.4/install.html}).
It is of crucial importance to check out the version tag 1.4.2 before
compiling and installing \TRIQS. This can be done by running the following
shell command

\begin{verbatim}
$ git checkout 1.4.2
\end{verbatim}

in the cloned source directory of the \TRIQS library.
It is also important to make sure that \TRIQS is compiled against a recent
version of the Boost C++ libraries, 1.58 or newer.

Installing \SOM is similar to that of other \TRIQS-based applications.
Assuming that \TRIQS 1.4.2 has been installed at \verb|/path/to/TRIQS/install/dir|
\SOM is simply installed by issuing the following commands at the shell prompt:

\begin{verbatim}
$ git clone https://github.com/krivenko/som som_src
$ mkdir som_build && cd som_build
$ cmake -DTRIQS_PATH=/path/to/TRIQS/install/dir ../som_src
$ make
$ make test
$ make install
\end{verbatim}
This will install \SOM in the same location as \TRIQS. Further installation
instructions are given in the online documentation.

\subsection{\label{sec:citation}Citation policy}

\TRIQS/\SOM is provided free of charge. We kindly ask to help us with its continued
maintenance and development by citing the present paper, as well as the
accompanying paper to the \TRIQS library \cite{TRIQS}, in any published work 
using the \SOM package. We also strongly recommend to cite 
\cite{Mishchenko2000} as the original work where the Stochastic Optimization 
Method was first introduced.

\subsection{\label{sec:contributing}Contributing}

\TRIQS/\SOM is an open source project and we encourage feedback and contributions from
the user community. Issues should be reported via the GitHub
website at \url{https://github.com/krivenko/som/issues}. For contributions,
we recommend to use the pull requests on the GitHub website. It is recommended 
to coordinate with the \TRIQS/\SOM developers before any major contribution.

\section{\label{sec:summary}Summary}

We have presented the open-source \TRIQS/\SOM package that implements the Stochastic
Optimization Method for analytic continuation. On a set of practical examples,
we demonstrated the versatility of \TRIQS/\SOM, as well as its potential to become a
viable alternative to various implementations of the Maximum Entropy Method,
the current {\it de facto} standard in the field of computational condensed
matter physics. The algorithm contains a number of performance optimizations
that significantly reduce overall computation costs.

Our plans for future releases include porting the code base to more recent
versions of the \TRIQS library and adding support for more integral kernels,
such as kernels for superconducting correlators \cite{Levy2017}. Another
interesting addition would be the ability to compute the two-point correlators
of particular solutions $\sigma_{mm'}$ as defined in \cite{Goulko2017}.
Following the methodology of Goulko {\it et al}, one could use the accumulated
correlators to characterize the accuracy of the output spectral functions.

\section{\label{sec:acknowledgements}Acknowledgements}

Authors are grateful to Alexander Lichtenstein and Olga Goulko for fruitful
discussions, to Roberto Mozara for his valuable comments on the
usability of the code, and to Andrey Mishchenko for giving access to the
original FORTRAN implementation of the SOM algorithm. We would also like to
thank Vladislav Pokorn\'y and Hanna Terletska, who stimulated writing of this
paper. I.K. acknowledges support from Deutschen Forschungsgemeinschaft (DFG) via
project SFB668 (subproject A3, ``Electronic structure and magnetism of
correlated systems''). M.H. acknowledges financial support by the DFG
in the framework of the SFB925. The computations were performed with
resources provided by the North-German Supercomputing Alliance (HLRN).

\appendix
\gdef\thesection{\Alph{section}} 
\makeatletter
\renewcommand\@seccntformat[1]{Appendix \csname the#1\endcsname.\hspace{0.5em}}
\makeatother

\section{\label{app:parameters}Advanced simulation parameters}

\begin{table}
    \centering
    \resizebox{\columnwidth}{!}{%
    \begin{tabularx}{\textwidth}{|c|c|c|X|}
        \hline
        Parameter Name & Python type & Default value & \makecell{Meaning} \\
        \hline\hline
        \py{random_seed} & \py{int} & \py{34788+928374*MPI.rank} &
        Seed for pseudo-random number generator.\\
        \hline
        \py{random_name} & \py{str} & \makecell{\py{"mt19937"}\\
            (Mersenne Twister\\19937 generator)} &
        Name of pseudo-random number generator. Other supported values are
        \py{mt11213b}, \py{ranlux3} and variants of
        \py{lagged_fibonacci} PRNG (see documentation of Boost.Random for
        more details).\\
        \hline
        \py{max_rects} & \py{int} & \py{60} & Maximum number of rectangles to 
        represent a configuration ($K$).\\
        \hline
        \py{min_rect_width} & \py{float} & \py{1e-3} &
        Minimal allowed width of a rectangle, in units of the energy window
        width.\\
        \hline
        \py{min_rect_weight} & \py{float} & \py{1e-3} &
        Minimal allowed weight of a rectangle, in units of the requested 
        solution norm $\mathcal{N}$.\\
        \hline
        \py{distrib_d_max} & \py{float} & \py{2.0} &
        Maximal parameter of the power-law distribution function for the 
        Metropolis algorithm ($d_\text{max}$).\\
        \hline
        \py{gamma} & \py{float} & \py{2.0} &
        Proposal probability parameter $\gamma$ (see 
        Appendix~\ref{app:prop_prob}).\\
        \hline
        \py{adjust_l_good_d} & \py{float} & \py{2.0} &
        Maximal ratio $D/D_\text{min}$ for a particular solution to be
        considered good, $\alpha_\text{good}$ (see Section~\ref{sec:method}).\\
        \hline
        \py{hist_max} & \py{float} & \py{2.0} &
        Right boundary of the histograms, in units of $D_\text{min}$
        (left boundary is always set to $D_\text{min}$).\\
        \hline
        \py{hist_n_bins} & \py{int} & \py{100} &
        Number of bins for the histograms.\\
        \hline
    \end{tabularx}
    }
    \caption{\label{tab:fine_parameters}Fine tuning \py{run()} parameters.}
\end{table}

\begin{table}
    \centering
    \resizebox{\columnwidth}{!}{%
    \begin{tabularx}{\textwidth}{|c|c|c|X|}
        \hline
        Parameter Name & Python type & Default value & \makecell{Meaning} \\
        \hline\hline
        \py{adjust_f_range} & \py{(int,int)} & \py{(100,5000)} &
        Search range for the $F$-adjustment procedure (see 
        Appendix~\ref{app:fit_quality}).\\
        \hline
        \py{adjust_f_l} & \py{int} & \py{20} &
        Number of particular solutions used in the $F$-adjustment procedure
        (see Appendix~\ref{app:fit_quality}).\\
        \hline
        \py{adjust_f_kappa} & \py{float} & \py{0.25} &
        Limiting value of $\kappa$ used in the $F$-adjustment procedure
        (see Appendix~\ref{app:fit_quality}).\\
        \hline
        \py{adjust_l_range} & \py{(int,int)} & \py{(100,2000)} &
        Search range for the $L$-adjustment procedure (see
        Section~\ref{sec:method}).\\
        \hline
        \py{adjust_l_verygood_d} & \py{float} & \py{4.0/3} &
        Maximal ratio $D/D_\text{min}$ for a particular solution to be
        considered very good, $\alpha_\text{very good}$(see 
        Section~\ref{sec:method}).\\
        \hline
        \py{adjust_l_ratio} & \py{float} & \py{0.95} &
        Critical ratio $L_\text{very good}/L_\text{good}$ that stops the
        $L$-adjustment procedure (see Section~\ref{sec:method}).\\
        \hline
    \end{tabularx}
    }
    \caption{\label{tab:adjust_parameters}\py{run()} parameters for $F$- and
        $L$-adjustment procedures (see Appendix~\ref{app:fit_quality} and
        Section~\ref{sec:method} respectively).}
\end{table}
\FloatBarrier

\section{\label{app:int_kernels_imfreq}Evaluation of integrated Matsubara
kernels}

\begin{table}
\begin{tabulary}{0.7\textwidth}{|c|c|}
    \hline
    Observable kind, $O$ &
    Integrated kernel,
    $\int_{\epsilon_\mathrm{min}}^{\epsilon_\mathrm{max}}
    d\epsilon\ K(\xi_n,\epsilon) R_{\{c,w,h\}}(\epsilon)$
    \rule{0pt}{2.6ex}\rule[-1.2ex]{0pt}{0pt}\\
    \hline\hline
    \makecell{Green's function\\of fermions} &
    $h \log\left(\frac{i\omega_n - c + w/2}{i\omega_n - c - w/2}\right)$
    \rule{0pt}{3.0ex}\rule[-1.6ex]{0pt}{0pt}
    \\\hline
    \makecell{Correlator of\\boson-like operators} &
    $\frac{hw}{\pi} + \frac{hi\nu_n}{\pi}\log
    \left(\frac{i\nu_n - c - w/2}{i\nu_n - c + w/2}\right)$
    \rule{0pt}{3.0ex}\rule[-1.6ex]{0pt}{0pt}
    \\\hline
    \makecell{Autocorrelator of a\\Hermitian operator} &
    $\frac{2hw}{\pi} + \frac{2h\nu_n}{\pi}\left(
    \atan\left(\frac{c-w/2}{\nu_n}\right) -
    \atan\left(\frac{c+w/2}{\nu_n}\right)
    \right)$
    \rule{0pt}{3.0ex}\rule[-1.6ex]{0pt}{0pt}
    \\\hline
    Zero temperature correlator &
    \makecell{$h \log\left(\frac{i z_n - c + w/2} {i z_n - c - w/2}\right)$, \\
    with $z_n = \pi(2n+1)/\tau_\text{max}$ or $2\pi n/\tau_\text{max}$}
    \rule{0pt}{5.2ex}\rule[-3.8ex]{0pt}{0pt}
    \\\hline
\end{tabulary}
    \caption{\label{tab:int_kernels_imfreq}Integrated kernels in the Matsubara representation.
    Respective kernels $K(\xi_n, \epsilon)$ are defined in the second column of Table \ref{tab:kernels}.
    $R_{\{c,w,h\}}(\epsilon)$ is the rectangle function defined in (\ref{eq:rectangle}).
    }
\end{table}
\FloatBarrier

\section{\label{app:int_kernels_imtime}Evaluation of integrated imaginary
time kernels}

The spline interpolation procedure used to evaluate the integrated kernels is
outlined in Sec.~\ref{sec:optimizations}. Here we only give the 
observable-specific details.
Throughout this Appendix, we denote dimensionless imaginary time and energy
arguments by $\alpha=\tau/\beta=1-\bar\alpha$ and $z = \Omega\beta$ 
respectively. $T_\alpha(z) = \int_0^z t_\alpha(x)dx$ denotes the integrated 
tail contribution. All series found in the tables below are rapidly 
(exponentially) convergent.
$\dilog(x)$ is the dilogarithm (Spence's function), $\psi(x) =
\frac{d}{dx}\log\Gamma(x)$ is the digamma function, $\psi^{(1)}(x) =
\frac{d^2}{dx^2}\log\Gamma(x)$ is the trigamma function, $\Psi(x) =
\frac{1}{2}[\psi(\frac{1+x}{2}) - \psi(\frac{x}{2})]$.

\newcommand{\auxsum}[1]{\ensuremath{\sum_{n=0}^\infty(-1)^n\frac{e^{#1z}}{#1}}}

\subsection{Green's function of fermions}

\begin{equation}
\Lambda(\tau;\Omega) = -\frac{1}{\beta}
\int\limits_0^{z}\frac{e^{-\alpha x}}{1+e^{-x}}dx =
-\frac{1}{\beta}
\left[\theta(-z)S^-_\alpha(z) + \theta(z)S^+_\alpha(z) + T_\alpha(z)\right],
\end{equation}
\begin{equation}
S^-_\alpha(z) = -\int\limits_z^0
\left[\frac{e^{-\alpha x}}{1+e^{-x}} - t_\alpha(x)
\right] dx,\quad
S^+_\alpha(z) = \int\limits_0^z
\left[\frac{e^{-\alpha x}}{1+e^{-x}} - t_\alpha(x)
\right] dx.
\end{equation}

Segments to construct the splines on:
\begin{equation}
S^-_\alpha(z): z\in[-z_0,0], \quad
S^+_\alpha(z): z\in[0;z_0], \quad
z_0 = -2\log(\mathtt{tolerance}).
\end{equation}

\begin{center}
    \resizebox{\columnwidth}{!}{%
    \begin{tabulary}{0.7\textwidth}{|c|c|c|c|c|}
    \hline
    $\alpha$ & $t_\alpha(x)$ & $T_\alpha(z)$ & 
    $S_\alpha^-(z)$ & 
    $S_\alpha^+(z)$\\
    \hline
    $0$ & $\theta(x)$ & $\theta(z)z$ & 
    $\log(1+e^{z})-\log(2)$ & $\log(1+e^{-z}) - \log(2)$ \\
    \hline
    $(0;1/2)$ & $\theta(x)e^{-\alpha x}$ & $\theta(z)\frac{e^{-\alpha 
    z}-1}{-\alpha}$ &
    $\begin{aligned}\auxsum{(n+\bar\alpha)}\\-\Psi(\bar\alpha)\end{aligned}$ &
    $\begin{aligned}-\auxsum{-(n+1+\alpha)}\\-\Psi(1+\alpha)\end{aligned}$ 
    \\
    \hline
    $1/2$ & $0$ & $0$ & $2\atan(e^{z/2})-\pi/2$ & $-2\atan(e^{-z/2})+\pi/2$ 
    \\
    \hline
    $(1/2;1)$ & $\theta(-x)e^{\bar\alpha x}$ & 
    $\theta(-z)\frac{e^{\bar\alpha z}-1}{\bar\alpha}$ & 
    $\begin{aligned}-\auxsum{(n+1+\bar\alpha)}\\+\Psi(1+\bar\alpha)\end{aligned}$
    &
    $\begin{aligned}\auxsum{-(n+\alpha)}\\+\Psi(\alpha)\end{aligned}$ \\
    \hline
    $1$ & $\theta(-x)$ & $\theta(-z)z$ & 
    $-\log(1+e^z) + \log(2)$ & $-\log(1+e^{-z}) + \log(2)$ \\
    \hline
    \end{tabulary}
    }
\end{center}

\subsection{Correlator of boson-like operators}

\begin{equation}
\Lambda(\tau;\Omega) = \frac{1}{\pi\beta^2}
\int\limits_0^{z}\frac{xe^{-\alpha x}}{1-e^{-x}}dx
= \frac{1}{\pi\beta^2}
\left[\theta(-z)S^-_\alpha(z) + \theta(z)S^+_\alpha(z) + T_\alpha(z)\right],
\end{equation}
\begin{equation}
S^-_\alpha(z) = -\int\limits_z^0
\left[\frac{xe^{-\alpha x}}{1-e^{-x}} - t_\alpha(x)
\right] dx,\quad
S^+_\alpha(z) = \int\limits_0^z
\left[\frac{xe^{-\alpha x}}{1-e^{-x}} - t_\alpha(x)
\right] dx.
\end{equation}

Segments to construct the splines on:
\begin{equation}
S^-_\alpha(z): z\in[-z_0,0], \quad
S^+_\alpha(z): z\in[0;z_0].
\end{equation}
Factor $x$ in the kernel tends to strengthen its delocalization, so it makes
sense to choose $z_0$ slightly larger, for instance,
$z_0=-2.3\log(\mathtt{tolerance})$.

\begin{center}
    \resizebox{\columnwidth}{!}{%
    \begin{tabulary}{0.7\textwidth}{|c|c|c|c|c|}
    \hline
    $\alpha$ & $t_\alpha(x)$ & $T_\alpha(z)$ & $S_\alpha^-(z)$ & 
    $S_\alpha^+(z)$\\
    \hline
    $0$ & $\theta(x)x$ & $\theta(z)z^2/2$ &
    $\begin{aligned}-\pi^2/6+z\log(1-e^z)\\+\dilog(e^z)\end{aligned}$ & 
    $\begin{aligned}\pi^2/6+z\log(1-e^{-z})\\-\dilog(e^{-z})\end{aligned}$ \\
    \hline
    $(0;1/2)$ & $\theta(x)xe^{-\alpha x}$ &
    $\theta(z)\frac{1-e^{-\alpha z}(1+z\alpha)}{\alpha^2}$ &
    $\begin{aligned}\sum_{n=0}^\infty\frac{e^{(n+\bar\alpha)z}[1-z(n+\bar\alpha)]}{(n+\bar\alpha)^2}
    \\-\psi^{(1)}(\bar\alpha)\end{aligned}$ &
    $\begin{aligned}-\sum_{n=1}^\infty\frac{e^{-(n+\alpha)z}[1+z(n+\alpha)]}{(n+\alpha)^2}
    \\+\psi^{(1)}(1+\alpha)\end{aligned}$ \\
    \hline
    $1/2$ & $0$ & $0$ & 
    $\begin{aligned}-\pi^2/2+z\log(\tanh(-z/4))\\-\dilog(e^z)+4\dilog(e^{z/2})\end{aligned}$
     &
    $\begin{aligned}\pi^2/2+z\log(\tanh(z/4))\\+\dilog(e^{-z})-4\dilog(e^{-z/2})\end{aligned}$
     \\
    \hline
    $(1/2;1)$ & $\theta(-x)(-x)e^{\bar\alpha x}$ &
    $\theta(-z)\frac{e^{\bar\alpha z}(1-z\bar\alpha)-1}{(\bar\alpha)^2}$ &
    $\begin{aligned}\sum_{n=1}^\infty\frac{e^{(n+\bar\alpha)z}[1-z(n+\bar\alpha)]}{(n+\bar\alpha)^2}
    \\-\psi^{(1)}(1+\bar\alpha)\end{aligned}$ & 
    $\begin{aligned}-\sum_{n=0}^\infty\frac{e^{-(n+\alpha)z}[1+z(n+\alpha)]}{(n+\alpha)^2}
    \\+\psi^{(1)}(\alpha)\end{aligned}$ \\
    \hline
    $1$ & $\theta(-x)(-x)$ & $\theta(-z)(-z^2/2)$ &
    $\begin{aligned}-\pi^2/6+z\log(1-e^z)\\+\dilog(e^z)\end{aligned}$ & 
    $\begin{aligned}\pi^2/6+z\log(1-e^{-z})\\-\dilog(e^{-z})\end{aligned}$ \\
    \hline
    \end{tabulary}
    }
\end{center}

\subsection{Autocorrelator of a Hermitian operator}

\begin{equation}
\Lambda(\tau;\Omega) = \frac{1}{\pi\beta^2}
\int\limits_0^{z}\frac{x(e^{-\alpha x}+e^{-\bar\alpha x})}
{1-e^{-x}}dx = \frac{1}{\pi\beta^2} \left[
S_\alpha(z) + T_\alpha(z) \right],
\end{equation}
\begin{equation}
S_\alpha(z) = \int\limits_0^z
\left[\frac{x(e^{-\alpha x}+e^{-(1-\alpha)x})}{1-e^{-x}} - t_\alpha(x)
\right] dx.
\end{equation}
(The upper integration limit can be only positive, which eliminates the need
for two splines). Segment to construct the spline on is $z\in[0;z_0]$, 
$z_0=-2.3\log(\mathtt{tolerance})$.

\begin{center}
    \resizebox{\columnwidth}{!}{%
    \begin{tabulary}{0.7\textwidth}{|c|c|c|c|}
        \hline
        $\alpha$ & $t_\alpha(x)$ & $T_\alpha(z)$ & $S_\alpha(z)$ \\
        \hline
        $0,1$ & $x(1+e^{-x})$ & $1+z^2/2-e^{-z}(1+z)$ &
        $\begin{aligned}-1+\pi^2/3+2z\log(1-e^{-z})\\
        -2\dilog(e^{-z})+e^{-z}(1+z)\end{aligned}$ \\
        \hline
        $(0;1/2)\cup(1/2;1)$ & $x(e^{-\alpha x}+e^{-\bar\alpha x})$ &
        $\begin{array}{c}\frac{1-e^{-\alpha z}(1+z\alpha)}{\alpha^2}\\+
        \frac{1-e^{-\bar\alpha z}(1+z\bar\alpha)}{(\bar\alpha)^2}\end{array}$ 
        &$\begin{aligned}-\sum_{n=1}^\infty\frac{e^{-(n+\alpha)z}[1+z(n+\alpha)]}
        {(n+\alpha)^2} +\psi^{(1)}(1+\alpha) \\
        -\sum_{n=1}^\infty\frac{e^{-(n+\bar\alpha)z}[1+z(n+\bar\alpha)]}
        {(n+\bar\alpha)^2} +\psi^{(1)}(1+\bar\alpha)\end{aligned}$ \\
        \hline
        $1/2$ & $2xe^{-x/2}$ & $4(2-e^{-z/2}(2+z))$ &
        $\begin{aligned}-8+\pi^2+4e^{-z/2}(2+z)+2z\log\tanh(z/4)\\
        -8\dilog(e^{-z/2})+2\dilog(e^{-z})\end{aligned}$ \\
        \hline
    \end{tabulary}
    }
\end{center}

\subsection{Zero temperature correlator}

\begin{equation}
\int\limits_{0}^{+\infty} d\epsilon\ K(\tau,\epsilon) R_{\{c,w,h\}}(\epsilon) = 
\begin{cases}
-hw, &\tau=0,\\
\frac{h}{\tau}(e^{-\tau(c+w/2)}-e^{-\tau(c-w/2)}) ,&\mathrm{otherwise}.
\end{cases}
\end{equation}

\section{\label{app:int_kernels_legendre}Evaluation of integrated Legendre
kernels}

The spline interpolation procedure used to evaluate the integrated kernels is
outlined in Sec.~\ref{sec:optimizations}.

\subsection{Green's function of fermions}

\begin{equation}
\Lambda(\ell;\Omega) = (-\mathrm{sgn}(\Omega))^{\ell+1}\sqrt{2\ell+1}
\int_0^{|\Omega|\beta/2} \frac{i_\ell(x)}{\cosh(x)} dx.
\end{equation}
Asymptotic form of the integrand ($x\to\infty$),
\begin{align}
\frac{i_\ell(x)}{\cosh(x)} &=
\frac{e^x}{e^x+e^{-x}}\sum_{n=0}^\ell(-1)^n
\frac{a_n(\ell+1/2)}{x^{n+1}} +
\frac{e^{-x}}{e^x+e^{-x}}\sum_{n=0}^\ell(-1)^{\ell+1}
\frac{a_n(\ell+1/2)}{x^{n+1}}\nonumber\\
&\approx
\sum_{n=0}^\ell(-1)^n \frac{a_n(\ell+1/2)}{x^{n+1}}.
\end{align}
Integral $F(z)=\int_0^z \frac{i_\ell(x)}{\cosh(x)} dx$ in the high-energy limit,
\begin{equation}
F^>(z)|_{z>x_0} = F^<(x_0) +
\left.\left\{
\log(x) +
\sum_{n=1}^\ell (-1)^{n+1}\frac{a_n(\ell+1/2)}{x^n n}
\right\}\right|_{x_0}^z.
\end{equation}

\subsection{Correlator of boson-like operators}

\begin{equation}
\Lambda(\ell;\Omega) = -(-\mathrm{sgn}(\Omega))^{\ell+1}
\frac{2\sqrt{2\ell+1}}{\pi\beta}
\int_0^{|\Omega|\beta/2} \frac{i_\ell(x) x}{\sinh(x)} dx.
\end{equation}
Asymptotic form of the integrand ($x\to\infty$),
\begin{align}
\frac{i_\ell(x) x}{\sinh(x)} &=
\frac{e^x}{e^x-e^{-x}}\sum_{n=0}^\ell(-1)^n
\frac{a_n(\ell+1/2)}{x^n} +
\frac{e^{-x}}{e^x-e^{-x}}\sum_{n=0}^\ell(-1)^{\ell+1}
\frac{a_n(\ell+1/2)}{x^n}\nonumber\\
&\approx
\sum_{n=0}^\ell(-1)^n \frac{a_n(\ell+1/2)}{x^n}.
\end{align}
Integral $F(z) = \int_0^z \frac{i_\ell(x)x}{\sinh(x)} dx$ in the high-energy
limit,
\begin{equation}
F^>(z)|_{z>x_0} = F^<(x_0) +
\left.\left\{
x - \frac{\ell(\ell+1)}{2}\log(x) +
\sum_{n=1}^{\ell-1} (-1)^n\frac{a_{n+1}(\ell+1/2)}{x^n n}
\right\}\right|_{x_0}^z.
\end{equation}

\subsection{Autocorrelator of a Hermitian operator}

\begin{equation}
\Lambda(\ell;\Omega) =
(1 + (-1)^\ell)\frac{2\sqrt{2\ell+1}}{\pi\beta}
\int_0^{\Omega\beta/2} \frac{i_\ell(x) x}{\sinh(x)} dx.
\end{equation}

This integral is evaluated in full analogy with the previous kernel.

\subsection{Zero temperature correlator}

\begin{equation}
\Lambda(\ell;\Omega) = (-1)^{\ell+1}\sqrt{2\ell+1}
\int_0^{\Omega\tau_{max}/2} 2e^{-x} i_\ell(x) dx.
\end{equation}
Asymptotic form of the integrand ($x\to\infty$),
\begin{equation}
2e^{-x} i_\ell(x) =
\sum_{n=0}^\ell(-1)^n
\frac{a_n(\ell+1/2)}{x^n} +
e^{-2x}\sum_{n=0}^\ell(-1)^{\ell+1}
\frac{a_n(\ell+1/2)}{x^n}
\approx\sum_{n=0}^\ell(-1)^n \frac{a_n(\ell+1/2)}{x^n}.
\end{equation}
Integral $F(z) = \int_0^z 2 e^{-x} i_\ell(x) dx$ in the high-energy limit,
\begin{equation}
F^>(z)|_{z>x_0} = F^<(x_0) + \left.\left\{
\log(x) + \sum_{n=1}^\ell (-1)^{n+1}\frac{a_n(\ell+1/2)}{x^n n}
\right\}\right|_{x_0}^z.
\end{equation}

\section{\label{app:prop_prob}Probability density function for the
parameter change}

Every proposed elementary update is parametrized by a real number
$\delta\xi \in[\delta\xi_\mathrm{min};\delta\xi_\mathrm{max}]$.
A concrete meaning of $\delta\xi$ depends on the elementary update in question.
For instance, $\delta\xi$ can be a shift of the centre of an existing 
rectangle, or the weight of a rectangle to be added. In general, $\delta\xi$ 
are defined so that larger $|\delta\xi|$ correspond to more prominent changes 
in the configuration. \SOM randomly generates values of $\delta\xi$ according to
the following probability density function,
\begin{equation}\label{eq:delta_xi_pdf}
\mathcal{P}(\delta\xi) = N
\exp\left(-\gamma \frac{|\delta\xi|}{X}\right), \quad
X \equiv \max(|\delta\xi_\mathrm{min}|, |\delta\xi_\mathrm{max}|),
\end{equation}
\begin{equation}
N = \frac{\gamma}{X}\left[
\sign(\delta\xi_\mathrm{min})(e^{-\gamma|\delta\xi_\mathrm{min}|/X} - 1) +
\sign(\delta\xi_\mathrm{max})(1 - e^{-\gamma|\delta\xi_\mathrm{max}|/X})
\right]^{-1}.
\end{equation}
User can change parameter $\gamma > 0$ to control non-uniformity of the
PDF~(\ref{eq:delta_xi_pdf}).

\section{\label{app:fit_quality}Fit quality criterion and choice of $F$}

Mishchenko introduced a special quantity $\kappa$ that characterizes the fit
quality of a given particular solution $A(\epsilon)$ \cite{Pavarini2012}.
\begin{equation}
\kappa = \frac{1}{M-1}\sum_{m=2}^M\theta(-\Delta(m)\Delta(m-1)).
\end{equation}
The observable $O$ and, therefore, deviation $\Delta(m)$ (Eq. (\ref{eq:objective_function}))
are assumed to be real-valued here. $\kappa$ measures the degree to what adjacent deviation
values $\Delta(m)$ are anti-correlated. Since input data points $O(\xi_m)$ are
expected to be statistically independent, deviations $\Delta(m)$ should
rapidly fluctuate changing sign between adjacent points. Conversely,
if the solution $A(\epsilon)$ gives a systematic deviation from the input points
$O(\xi_m)$, products $\Delta(m)\Delta(m-1)$ are more likely to be positive,
which results in a smaller $\kappa$.
Ideally, $\kappa$ must approach $1/2$, but in practice values as small as
$1/4$ signal satisfactory fit quality.

A similar expression can be written for complex-valued observables, such
as functions of Matsubara frequencies. Our generalization consists in replacing
\begin{multline}
\theta(-\Delta(m)\Delta(m-1)) \mapsto
\frac{1}{2}\left[1 - 
\frac{\Re[\Delta(m)\Delta^*(m-1)]}{|\Delta(m)\Delta^*(m-1)|} \right]\\=
\frac{1}{2}\left[ 1 - \cos[\arg(\Delta(m)) - \arg(\Delta(m-1))] \right].
\end{multline}
According to the modified definition, two adjacent values of $\Delta$ are considered anti-correlated, if the complex phase shift between them exceeds $\pm\pi/2$. In the case of real-valued quantities the phase shift is always either 0, or $\pi$.

\SOM has an option to automatically adjust the number of global updates $F$
used in a calculation. Starting from $F = F_\text{min}$ it finds a few (20 by
default) particular solutions $\tilde A(\epsilon)$ and checks that
$\kappa\geq\kappa_\text{min}$ (1/4 by default) for at least a half of them.
If this is not the case, the test is repeated with an increased $F$. This
procedure stops upon a successful $\kappa$-test or when $F_\text{max}$ is
reached, whichever happens first.

\bibliographystyle{elsarticle-num.bst}
\bibliography{main}

\end{document}